\documentclass[aps,prb,twocolumn,superscriptaddress,showpacs,amsmath,amssymb,
footinbib,longbibliography,10pt]{revtex4-2}
\usepackage[utf8]{inputenc}
\usepackage{bm}
\usepackage[usenames]{color}
\usepackage{multirow}
\usepackage{amssymb}
\usepackage{amsbsy}
\usepackage{amsmath}
\usepackage{stmaryrd}
\usepackage{graphicx}
\usepackage{epsfig}
\usepackage{placeins}
\usepackage[normalem]{ulem}
\usepackage{bbold}
\usepackage{bbm}
\usepackage{braket}
\usepackage{graphicx}
\usepackage{tikz}
\usepackage{xcolor}

\usetikzlibrary{calc}
\usetikzlibrary{patterns}
\usepackage[colorlinks,linkcolor=blue,citecolor=blue,urlcolor=blue]{hyperref}

\definecolor{mycolor}{RGB}{0,100,0} 
\makeatletter

\setcitestyle{numbers,square,comma,sort&compress}

\renewcommand{\eqref}[1]{Eq.\hspace{2pt}\protect(\ref{#1})}
\newcommand\e[1]{\mathrm{e}^{#1}} 

\definecolor{palette1}{HTML}{A8216B}
\definecolor{palette2}{HTML}{F1184C}
\definecolor{palette3}{HTML}{F36943}
\definecolor{palette4}{HTML}{F7DC66}
\definecolor{palette5}{HTML}{2E9599}

\usepackage{scalerel}
\usepackage{tikz}
\usetikzlibrary{calc}
\usetikzlibrary{patterns}
\usetikzlibrary{svg.path}
\definecolor{orcidlogocol}{HTML}{A6CE39}
\tikzset{
  orcidlogo/.pic={
    \fill[orcidlogocol]
svg{M256,128c0,70.7-57.3,128-128,128C57.3,256,0,198.7,0,128C0,57.3,57.3,0,128,
0C198.7,0,256,57.3,256,128z};
    \fill[white] svg{M86.3,186.2H70.9V79.1h15.4v48.4V186.2z}

svg{M108.9,79.1h41.6c39.6,0,57,28.3,57,53.6c0,27.5-21.5,53.6-56.8,53.6h-41.8V79.
1z
M124.3,172.4h24.5c34.9,0,42.9-26.5,42.9-39.7c0-21.5-13.7-39.7-43.7-39.7h-23.
7V172.4z}

svg{M88.7,56.8c0,5.5-4.5,10.1-10.1,10.1c-5.6,0-10.1-4.6-10.1-10.1c0-5.6,4.5-10.1
,10.1-10.1C84.2,46.7,88.7,51.3,88.7,56.8z};
  }
}

\newcommand\orcid[1]{\!%
  \href{https://orcid.org/#1}{%
    \mbox{%
      \scaleto{%
        \begin{tikzpicture}[yscale=-1,transform shape]
          \pic{orcidlogo};
        \end{tikzpicture}
      }{8pt}%
    }%
  }%
}

\begin{document}
\title{Nontrivial damping of magnetization currents in perturbed spin chains}

\author{Mariel Kempa~\orcid{0009-0006-0862-4223}}
\email{makempa@uos.de}
\affiliation{University of Osnabr{\"u}ck, Department of Mathematics/Computer
Science/Physics, D-49076 Osnabr{\"u}ck, Germany}

\author{Markus Kraft~\orcid{0009-0008-4711-5549}}
\affiliation{University of Osnabr{\"u}ck, Department of Mathematics/Computer
Science/Physics, D-49076 Osnabr{\"u}ck, Germany}

\author{Jiaozi Wang~\orcid{0000-0001-6308-1950}}
\affiliation{University of Osnabr{\"u}ck, Department of Mathematics/Computer
Science/Physics, D-49076 Osnabr{\"u}ck, Germany}

\author{Robin Steinigeweg~\orcid{0000-0003-0608-0884}}
\email{rsteinig@uos.de}
\affiliation{University of Osnabr{\"u}ck, Department of Mathematics/Computer
Science/Physics, D-49076 Osnabr{\"u}ck, Germany}

\date{\today}


\begin{abstract}
Since perturbations are omnipresent in physics, understanding their impact on
the dynamics of quantum many-body systems is a vitally important but
notoriously difficult question. On the one hand, random-matrix and typicality
arguments suggest a rather simple damping in the overwhelming majority of cases,
e.g., exponential damping according to Fermi's Golden Rule. On the other hand,
counterexamples are known to exist, and it remains unclear how frequent and
under which conditions such counterexamples appear. In our work, we consider the
spin-1/2 XXZ chain as a paradigmatic example of a quantum many-body system and
study the dynamics of the magnetization current in the easy-axis regime. Using
numerical simulations based on dynamical quantum typicality, we show that the
standard autocorrelation function is damped in a nontrivial way and that
only a modified version of this function is damped in a simple manner.
Employing projection-operator techniques in addition, we demonstrate that both,
the nontrivial and simple damping relation can be understood on perturbative
grounds. Our results are in agreement with earlier findings for the particle
current in the Hubbard chain.
\end{abstract}

\maketitle


\section{Introduction}

Quantum many-body systems out of equilibrium continue to be a central
topic of modern research and are relevant to a broad class of physical
situations. These situations range from strictly isolated quantum systems to
open or driven quantum systems explicitly coupled to external baths or forces.
In particular, substantial progress has been made in recent years
\cite{polkovnikov2011, eisert2015, bloch2008, dalessio2016, borgonovi2016,
abanin2019, bertini2021, landi2022, schollwoeck2005, schollwoeck2011,
weimer2021}, not least due to the advance of controlled experimental platforms,
the development of powerful numerical techniques, and the invention of fresh
theoretical concepts.

Within the diverse questions, a key issue is the equilibration and
thermalization of closed quantum systems. Here, the fundamental question arises
whether a system will eventually reach thermal equilibrium. A main ansatz is
the eigenstate thermalization hypothesis (ETH) \cite{deutsch1991,
srednicki1994, rigol2008}, which provides a microscopic explanation for the
emergence of thermalization and is closely related to random-matrix theory.
Moreover, equally important questions are the actual time scales of
equilibration and the specific form of the relaxation process
\cite{goldstein2013, reimann2016, garciapintos2017, richter2019, alhambra2020,
lezama2021, hamazaki2022, bartsch2024, wang2024, eckseler2025}.

In this context, an intriguing question is how the relaxation process is
altered if the system's Hamiltonian $H$ is affected by the presence of a
perturbation $V$ of some strength $\epsilon$ \cite{zotos2004, jung2006,
jung2007, steinigeweg2016, denardis2021, mallayya2021, roy2023,
nandy2023, gopalakrishnan2024, chen2025},
\begin{equation}
H = H_0 +\epsilon V \, .
\end{equation}
In general, this question is notoriously difficult to answer and the effect
of such a perturbation can be manifold. However, random-matrix theory and
typicality
arguments suggest a rather simple damping in the vast majority of cases
\cite{dabelow2020, dabelow2021, richter2020}, e.g., exponential according to
Fermi's Golden Rule,
\begin{equation}\label{eq:Golden rule}
f_\varepsilon(t) = f_{\varepsilon=0}(t) \, e^{-\varepsilon^2 \gamma t} \ ,
\end{equation}
which also includes scenarios, where already the unperturbed dynamics posses
rich features \cite{karrasch2014}. Still, counterexamples are known to exist
\cite{heitmann2021}, and it remains unclear how frequent and under which
conditions such counterexamples appear \cite{lamann2022}.

In our work, we discuss this impact of perturbations on the dynamics of quantum
many-body systems, with the focus on the magnetization current in the
spin-$1/2$ XXZ chain, serving as a paradigmatic example for observable and model
\cite{bertini2021}. In particular, we study the damping of the current
autocorrelation function for two distinct choices of reference systems, where
the first one is noninteracting \cite{steinigeweg2010, steinigeweg2011} and
second one is interacting \cite{denardis2022, prelovsek2022, kraft2024,
prelovsek2025}.
To this end, we use numerical simulations based on the concept of dynamical
quantum typicality
\cite{hams2000, gemmer2003, goldstein2006, popescu2006, reimann2007,
bartsch2009, white2009, bartsch2011, sugiura2012, elsayed2013, steinigeweg2014,
steinigeweg2015, heitmann2020, jin2021, mitric2024}, in order to obtain exact
results for systems of
comparatively large but still finite size. We complement these results by
further results from a lowest-order perturbation theory on the basis of the
time-convolutionless (TCL) projection-operator technique
\cite{chaturvedi1979, breuer2007}.

For the noninteracting reference system, we find an exponential damping, which
is consistent with other works. For the interacting
reference system, however, we show that the standard autocorrelation function
is damped in a nontrivial way, that does not follow Eq.\ (\ref{eq:Golden
rule}), and only a modified version of it is damped in a simple manner, that
follows Eq.\ (\ref{eq:Golden rule}). This modified version can be seen
as a correlation function in the interaction picture instead of the
Schr\"odinger picture, as defined later in Eq.\ (\ref{eq:mod_corr}).
We demonstrate that both, the nontrivial and simple damping can be also understood
on perturbative grounds. Our results are in agreement with earlier findings for
the particle current in the Hubbard chain \cite{heitmann2021}, and they suggest
that nontrivial damping occurs if the unperturbed system already has a rich
dynamical behavior.

This paper is structured as follows. In Sec.\ \ref{sec:model_and_observable}, we
first introduce the models studied, i.e., the integrable spin-$1/2$ XXZ chain
and its nonintegrable version with next-nearest-neighbor interactions.
Additionally, we define the perturbation scenarios considered as well as the
magnetization current as the observable of interest. In this context, we also
introduce the standard autocorrelation function in the limit of high
temperatures. In Sec.\ \ref{sec:TCL}, we turn to the TCL projection-operator
technique as our perturbative approach to the relaxation process, focusing on
the general framework in Sec.\ \ref{sec:TCL_framework}. In Sec.\
\ref{sec:TCL_simple_projection_and_modified_correlation_function}, a specific
choice of a projection is discussed, which has been used in previous works to
access a modified version of the standard current autocorrelation function.
A main extension of our work is presented in Sec.\
\ref{sec:TCL_extended_projection_and_standard_correlation_function}, where we
introduce another projection to connect to the desired standard current
autocorrelation function. Then, we turn to our results in Sec.\
\ref{sec:results} and provide a comparison between exact data and our TCL
results. We present our results for the noninteracting reference system in
Sec.\ \ref{sec:results_noninteracting} and for the interacting reference
system in Sec.\ \ref{sec:results_interacting}. In Sec.\ \ref{sec:conclusions},
we summarize our work and draw conclusions.


\section{Model and observable} \label{sec:model_and_observable}

\begin{figure}[t]
\includegraphics[width=0.8\columnwidth]{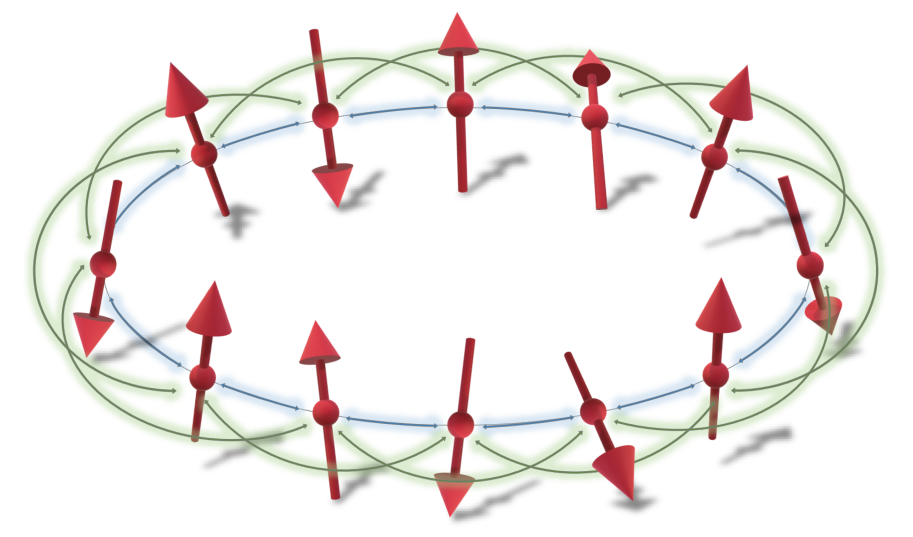}
\caption{Sketch of the spin chain, as defined in Eq.\ (\ref{eq:model}), with
periodic boundary conditions and interactions between both, nearest (blue) and
next-nearest (green) neighbors.}
\label{fig:sketch}
\end{figure}

Although our main question is of relevance to a broad class of quantum
many-body systems, we focus here on the spin-$1/2$ XXZ chain, which is a
paradigmatic example and has attracted significant attention in the literature
before \cite{bertini2021}. This model is given by the Hamiltonian
\begin{equation}\label{eq: integrable system}
H_\text{XXZ} = J \sum_{r=1}^{L} (S_{r}^{x} S_{r+1}^{x} + S_{r}^{y} S_{r+1}^{y}
+ \Delta S_{r}^{z} S_{r+1}^{z} ) \, ,
\end{equation}
where $S_r^i$ ($i = x$, $y$, $z$) are the components of a spin-$1/2$ operator
at lattice site $r$, $L$ is the total number of lattice sites, $J$ is the
exchange coupling constant, and $\Delta$ is the anisotropy in $z$ direction.
Throughout our work, we use periodic boundary conditions,
i.e., $S_{L+1}^i = S_1^i$.
Since the spin-$1/2$ XXZ chain is Bethe-Ansatz integrable for any value of the
anisotropy $\Delta$, we break this integrability by an additional interaction
between next-nearest sites,
\begin{equation} \label{eq:model}
H = H_\text{XXZ} + J \Delta' \sum_{r=1}^{L} S_{r}^{z} S_{r+2}^{z} \, ,
\end{equation}
where $\Delta'$ is the interaction strength. As will be discussed later, this
interaction will play the role of a perturbation.
The model is sketched in
Fig.\ \ref{fig:sketch}.

For all choices of $\Delta$ and $\Delta'$, the total magnetization
$S^z = \sum_r S^z_r$ is conserved, $[S^z, H] = 0$. 
As a consequence, the
temporal change of a local magnetization is related to a flow of local currents,
\begin{equation}
\frac{\text{d}}{\text{d}t} S_r^z = i [H, S_r^z] = j_{r-1}-j_r \, ,
\end{equation}
which then leads to the well-known definition of the total current
\cite{bertini2021}

\begin{equation}
j = \sum_{r=1}^{L} j_r \, , \quad j_r = J (S_{r}^{x} S_{r+1}^{y} - S_{r}^{y}
S_{r+1}^{x}) \, .
\end{equation}

While the total current $j$ itself does not depend on $\Delta$ or
$\Delta'$, its dynamics does. In particular, only in the case $\Delta = \Delta'=
0$, $j$ is conserved, $[j, H] = 0$. In general, one has
$\text{tr}[j] = 0$ and $\text{tr}[j^2] = L d J^2/8$, where $d = 2^L$ is the
dimension of the Hilbert space.

Within the framework of linear response theory \cite{kubo1991}, transport
properties are related to the autocorrelation function of $j$ at equilibrium,
\begin{equation}
\langle j(t) j(0) \rangle _{\text{eq}} = \frac{\text{tr}[\e{-\beta H} \e{i
H t} j \e{-i H t} j]}{Z} \, ,
\end{equation}
with the partition function $Z = \text{tr}[\e{-\beta H}]$ at the inverse
temperature
$\beta = 1/T$. 
For the remainder of our work, we focus on the limit of high
temperatures,
\begin{equation} \label{eq:standard}
C_S(t) = \lim_{\beta \to 0} \langle j(t) j(0) \rangle _{\text{eq}} =
\frac{\text{tr}[\e{i H t} j \e{-i H t} j]}{d} \, ,
\end{equation}
where $d$ denotes again the Hilbert-space dimension. 
Even for high temperatures, the time evolution of the
current autocorrelation function $C_S(t)$ remains nontrivial, has been
extensively studied in the literature \cite{bertini2021}, and can lead to
qualitatively different types of transport. Specifically, for the integrable
case $\Delta^\prime = 0$, magnetization transport is ballistic for $\Delta <
1$, superdiffusive for $\Delta = 1$, and diffusive for $\Delta > 1$.
Here, we focus on the diffusive regime and choose large anisotropies $\Delta
\gg 1$, where $C_S(t)$ features rich dynamics \cite{karrasch2014} until it
eventually decays to zero.

Note that an index $S$ is used to indicate a standard
correlation function and another index $M$ is used later to
indicate a modified correlation function, which can be seen
as a correlation function in the interaction picture instead of the
Schr\"odinger picture, as defined later in Eq.\ (\ref{eq:mod_corr}).


\begin{figure}[t]
\includegraphics[width=0.9\columnwidth]{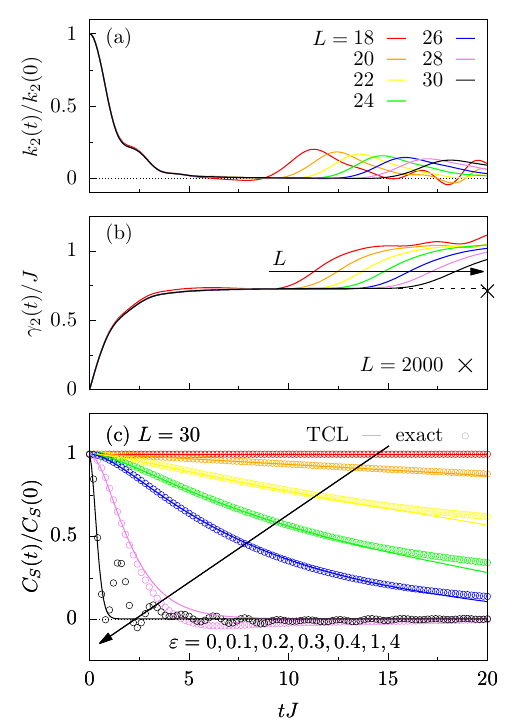}
\caption{Noninteracting reference system $H_0$ with $\Delta = 0$ and $\Delta' =
0$. (a) Second-order kernel $k_2(t)$ and (b) corresponding rate $\gamma_2(t)$,
both for various system sizes $L \leq 30$. (c) Resulting prediction for the
current autocorrelation function $C_S(t)$ and exact numerical results for
various perturbation strengths $\varepsilon$, both for a fixed system size
$L=30$. In (b), we also depict a $L=2000$ data point from Ref.\
\cite{steinigeweg2011}.}
\label{fig:Fig1}
\end{figure}

Since we are interested in the impact of a perturbation on the dynamics of
the autocorrelation function $C_S(t)$, we express the full Hamiltonian as
\begin{equation} \label{eq:scenario}
H = H_0 + \varepsilon V \, ,
\end{equation}
where $H_0$ is the unperturbed Hamiltonian and $\varepsilon V$ is a
perturbation of strength $\varepsilon$, which can be weak but also strong.
Specifically, we consider two different scenarios in our work. (i) The
reference Hamiltonian
\begin{equation}
H_0 = H(\Delta = \Delta ' = 0)
\end{equation}
is noninteracting and the perturbation
\begin{equation}
V = J \sum_{r=1}^{L} (S_{r}^{z} S_{r+1}^{z} + S_{r}^{z} S_{r+2}^{z}) \, , \quad
\varepsilon = \Delta = \Delta'
\end{equation}
contains all interactions. (ii) The reference Hamiltonian is changed into
\begin{equation}
H_0 = H(\Delta \neq 0, \Delta ' = 0)
\end{equation}
and also includes interactions between nearest sites, while the perturbation
\begin{equation}
V = J \sum_{r=1}^{L} S_{r}^{z} S_{r+2}^{z}\ , \quad \varepsilon = \Delta'
\end{equation}
consists of the other interactions between next-nearest sites. In the
scenario (i), the total current $j$ is conserved in the unperturbed system
$H_0$, $[j, H_0] = 0$. Hence, the autocorrelation function becomes
\begin{equation}
C_S(t) = \text{const.}
\end{equation}
for $\varepsilon = 0$ and has no dynamics at all.
In contrast, in the scenario
(ii), $[j, H_0] \neq 0$. Therefore, the autocorrelation function becomes
\begin{equation}
C_S(t) \neq \text{const.}
\end{equation}
for $\varepsilon = 0$ and has a dynamics, which can be also rich for large
anisotropies $\Delta \gg 1$ \cite{karrasch2014}.

For both scenarios, we study the change of $C_S(t)$ with $\varepsilon$, using
numerical simulations based on the concept of dynamical quantum typicality
\cite{heitmann2020, jin2021}. This concept allows us to treat finite systems of
sizes up to $L = 30$, where the Hilbert-space dimension is $D = {\cal
O}(10^9)$. These exact numerical simulations are complemented by perturbative
approaches, which will be introduced below.


\section{Projection-operator techniques} \label{sec:TCL}

\subsection{Framework}\label{sec:TCL_framework}

Next, we turn to our perturbative approaches, which rely  on
projection-operator techniques. These techniques can be applied to the
scenario in Eq.\ (\ref{eq:scenario}), where the full Hamiltonian $H$ is
decomposed into a reference system $H_0$ and a perturbation $V$.

The cental idea of projection-operator techniques is to reduce the complete
microscopic dynamics to equations of motion for a set of relevant variables,
which should at least include the observable of interest. To this end, one
formally defines a projection (super)operator
\begin{equation} \label{eq:projection}
\mathcal{P} \rho(t)= \sum_{i=1}^N \frac{\text{tr}[ P_i \, \rho(t) ]}
{\text{tr}[ (P_i)^2 ]} P_i \, ,
\end{equation}
where $\rho(t)$ is the time-dependent density matrix and the operators $P_i$
are orthogonal to each other,
\begin{equation}
\text{tr}[ P_i P_j ] = \text{tr}[ (P_i)^2] \, \delta_{i,j} \, .
\end{equation}
Therefore, ${\cal P}^2 = {\cal P}$. Throughout our work, we consider initial
states
\begin{equation}
\rho(0) = \text{span}[P_1, P_2, \ldots, P_N]
\end{equation}
such that $(1 - {\cal P}) \rho(0) = 0$.

Once the decomposition in Eq.\ (\ref{eq:scenario}) and the projection operator
in Eq.\ (\ref{eq:projection}) are chosen, the time-convolutionless (TCL)
projection-operator technique \cite{chaturvedi1979, breuer2007} then leads
to a time-local differential equation of the form
\begin{equation}\label{eq:TCL}
\frac{\text{d}}{\text{d} t} \mathcal{P} \rho_\text{I} (t) =
\mathcal{G}(t) \mathcal{P} \rho_\text{I} (t) + \mathcal{I} (t) (1-\mathcal{P})
\rho (0)
\end{equation}
and avoids the often troublesome time convolution in the context of the
Nakajima-Zwanzig formalism. Here, $\rho_\text{I}(t)$ is the time-dependent
density matrix in the interaction picture,
\begin{equation}
\rho_\text{I}(t) = e^{iH_0 t} e^{-iHt} \rho(0) e^{iHt} e^{-iH_0t} \, ,
\end{equation}
and the inhomogeneity ${\cal I}(t)$ can be neglected, due to our choice of
initial conditions. The generator $\mathcal{G}(t)$ is given as a systematic
perturbation expansion in powers of the perturbation strength $\varepsilon$,
\begin{equation}
{\cal G}(t) = \sum_{i=1}^\infty \varepsilon^i \, {\cal G}_i(t) \, , \quad {\cal
G}_{2i-1}(t) = 0 \, ,
\end{equation}
where odd contributions of the expansion vanish in many cases, as for our
choice below. Hence, the lowest order is the second order and reads
\begin{equation}
\mathcal{G}_2(t) = \varepsilon^2 \int_0^t \mathrm{d}t' \,
\mathcal{P} \mathcal{L}(t) \mathcal{L}(t') \mathcal{P} \, ,
\end{equation}
where the Liouvillian is given by
\begin{equation}
\mathcal{L}(t) \bullet = -i [V_\text{I}(t), \bullet] \, , \quad V_\text{I}(t) =
e^{iH_0 t} V(0) e^{-iH_0t} \, .
\end{equation}
Our central assumption is a corresponding truncation to this second order.
Thus, we rely on
\begin{equation} \label{eq:TCL2}
\frac{\text{d}}{\text{d} t} \mathcal{P} \rho_\text{I} (t) =
\mathcal{G}_2(t) \mathcal{P} \rho_\text{I} (t) \, .
\end{equation}
In general, the quality of such a truncation depends on the strength of
$\varepsilon$, the structure of $V$, and the choice of ${\cal
P}$.
Naturally, the quality can be also improved by taking into account higher-order
corrections. For the purposes of our work, however, the second-order truncation
will be sufficient.

In the following, we elaborate on the lowest-order TCL description in
Eq.\ (\ref{eq:TCL2}). We do so for two different choices of the projection
operator. The first projection is based on a previous work \cite{heitmann2021}
and yields to a prediction for a modified autocorrelation function $C_M(t)$,
which differs from the standard autocorrelation function $C_S(t)$. The second
projection aims at a perturbative description of the actual $C_S(t)$.


\subsection{Simple projection and modified correlation function}
\label{sec:TCL_simple_projection_and_modified_correlation_function}

As done in a previous work \cite{heitmann2021}, we first choose for the
projection only two operators, i.e., the identity $1$ and the total current
$j$. Thus, the projection reads
\begin{equation} \label{eq:simple_projection}
\mathcal{P} \rho_\text{I}(t) = \frac{1}{d} + \frac{C_M(t)}{\text{tr}[ j^2 ]} \,
j \, , \quad C_M(t) = \text{tr}[j \rho_\text{I}(t) ] \, ,
\end{equation}
where $C_M(t)$ is the time-dependent part of the projected density matrix. For
initial states
\begin{equation} \label{eq:initial_condition}
\rho(0) = \frac{1 + \alpha j}{d} \, ,
\end{equation}
this time-dependent part becomes
\begin{equation} \label{eq:mod_corr}
C_M(t) = \alpha  \, \frac{\text{tr}[\e{i H t} j \e{-i H t} \e{i H_0 t} j \e{-i
H_0 t}]}{d}
\end{equation}
and then takes on the form of a current autocorrelation function. Yet,
compared to the standard one in Eq.\ (\ref{eq:standard}), it is modified:
While one $j$ evolves w.r.t.\ $H$, the other $j$ evolves w.r.t.\ $H_0$. To
indicate this difference, we use the index $M$.

\begin{figure}[b]
\includegraphics[width=0.9\columnwidth]{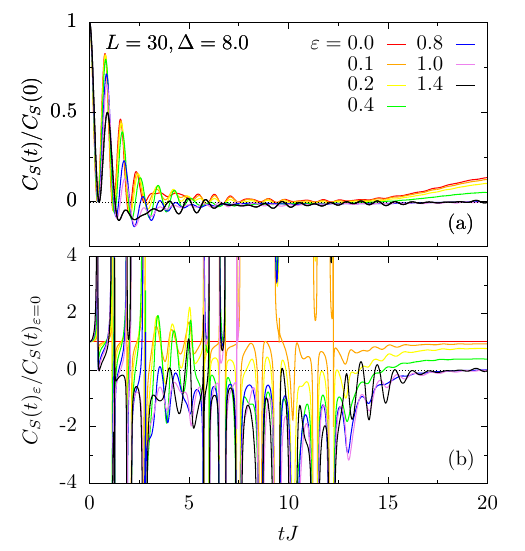}
\caption{Interacting reference system $H_0$ with $\Delta = 8.0$ and $\Delta' =
0$. (a) Exact dynamics of the autocorrelation function $C_S(t)$
for various perturbation strengths $\varepsilon$ and fixed system size $L=30$.
(b) Ratio between perturbed and unperturbed autocorrelation functions.}
\label{fig:Fig2}
\end{figure}

For the projection in Eq.\ (\ref{eq:simple_projection}), the second-order TCL
description in Eq.\ (\ref{eq:TCL2}) leads to the rate equation
\begin{equation} \label{eq:TCL2_first}
\frac{\text{d}}{\text{d} t} \, C_M(t) = -\varepsilon^2 \, \gamma_2(t) \, C_M(t)
\, ,
\end{equation}
where the time-dependent rate
\begin{equation}
\gamma_2 (t) = \int_0^t \text{d}t' \, k_2(t, t')
\end{equation}
is given as an time integral over the kernel
\begin{equation} \label{eq:k2}
k_2(t, t') = \frac{\text{tr}(i [j, V_I(t)] \, i[j, V_I(t')] )}{
\text{tr}( j^2 )} \, .
\end{equation}
As an additional approximation we use $k_2(t, t') =
k_2(t-t')$, which holds strictly only if $[j,H_0] = 0$. (See Appendix
\ref{sec:reduction} for further discussion and numerical verification.)
Obviously, Eq.\ (\ref{eq:TCL2_first}) is solved via an exponential decay of
the form
\begin{equation} \label{eq:solution}
\frac{C_M(t)}{C_M(0)} = \exp \Big[ -\varepsilon^2 \! \int_0^t \text{d}t' \,
\gamma_2(t') \Big ] \, ,
\end{equation}
with a generally time-dependent rate $\gamma_2(t)$. However, if the underlying
kernel $k_2(t-t')$ decays to zero on a certain time scale, $\gamma_2(t)$ becomes
constant after this time scale. Then, the earlier time dependence of
$\gamma_2(t)$ can be neglected in the case of a small perturbation strength
$\varepsilon$ and a slow relaxation process.

Overall, as a consequence of the solution in Eq.\ (\ref{eq:solution}), one
obtains a simple damping relation for the modified current autocorrelation
function,
\begin{equation}
\frac{C_M(t)_{\varepsilon}}{C_M(t)_{\varepsilon = 0}} = \exp \Big[
-\varepsilon^2 \! \int_0^t \text{d}t' \,
\gamma_2(t') \Big ] \, .
\end{equation}
It is important to note that, by
construction, this relation does not necessarily carry over to the standard
current autocorrelation function,
\begin{equation}
\frac{C_S(t)_{\varepsilon}}{C_S(t)_{\varepsilon = 0}}
\neq \frac{C_M(t)_{\varepsilon}}{C_M(t)_{\varepsilon = 0}} \, ,
\end{equation}
because both functions are only identical for $[j, H_0] = 0$.
This fact has been demonstrated in numerical simulations of
the one-dimensional Hubbard model~\cite{heitmann2021}, where $C_S(t)$ has
a complex damping, that does not follow
Eq.~(\ref{eq:Golden rule}), in contrast to $C_M(t)$.
Later, we will demonstrate a similar behavior for the spin-$1/2$ XXZ chain.
\begin{figure}[t]
\includegraphics[width=0.9\columnwidth]{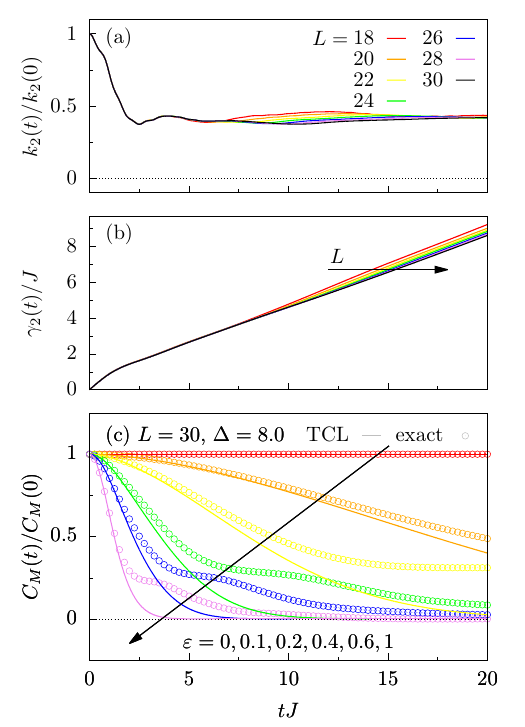}
\caption{Interacting reference system $H_0$ with $\Delta = 8.0$ and $\Delta' =
0$. (a) Second-order kernel $k_2(t)$ and (b) corresponding rate $\gamma_2(t)$,
both for various system sizes $L \leq 30$. (c) Resulting prediction for the
modified current autocorrelation function $C_M(t)$ and exact numerical results
for various perturbation strengths $\varepsilon$, both for a fixed system size
$L=30$.}
\label{fig:Fig3}
\end{figure}

\begin{figure}[b]
\includegraphics[width=0.9\columnwidth]{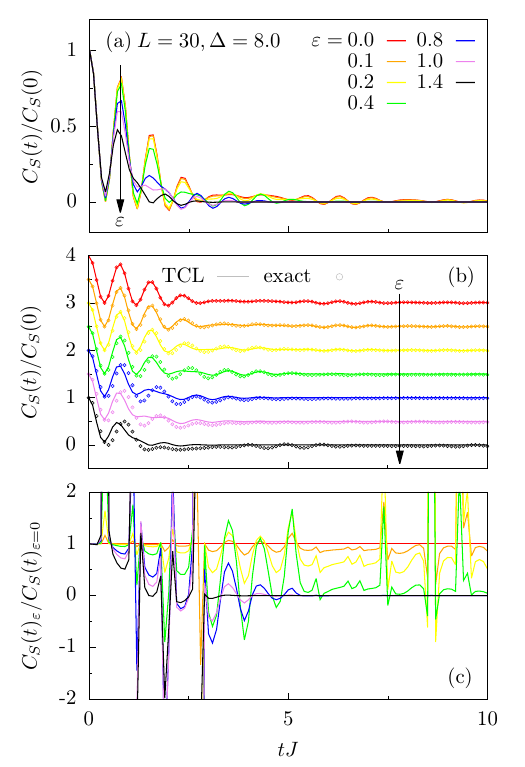}
\caption{Interacting reference system $H_0$ with $\Delta = 8.0$ and $\Delta' =
0$. (a) Second-order prediction for the dynamics of the standard current
autocorrelation function $C_S(t)$ for various perturbation strengths
$\varepsilon$ and a fixed system size $L=30$. (b) Comparison of (a) with the
exact dynamics, where data is shifted for better visibility. (c) Ratio between
perturbed and unperturbed
time evolution in (a).}
\label{fig:Fig4}
\end{figure}

\subsection{Extended projection and standard correlation function}
\label{sec:TCL_extended_projection_and_standard_correlation_function}

So far, our choice of the projection $\mathcal{P}$ in Eq.\
(\ref{eq:simple_projection}) allows us to obtain a perturbative description
for the modified current autocorrelation function $C_M(t)$, but not for the
standard current autocorrelation function $C_S(t)$. Hence, we go beyond
previous works and change the projection in a suitable way. The basic idea is an
extension of the projection by a third operator $P_3$, while the already used
operators $P_1 = 1$ and $P_2 = j$ are kept. Specifically, we choose
\begin{equation}
P_3 = j_\tau = e^{i H_0 \tau} j e^{-i H_0 \tau} \, ,
\end{equation}
which is the time evolution of $j$ w.r.t.\ $H_0$. Here, $\tau$ is an arbitrary
but fixed parameter. While $j_\tau$ is orthogonal to $P_1 = 1$,
it is not orthogonal to $P_2 = j$. Therefore, we change the second operator
into
\begin{equation}
P_2 = j' = j - \frac{\text{tr}[j_\tau j]}{\text{tr}[(j_\tau)^2]} \, j_\tau \, .
\end{equation}
Thus, the extended projection reads
\begin{equation} \label{eq:complicated_projection}
\mathcal{P} \rho_\text{I}(t) = \frac{1}{d} + \frac{a(t)}{\text{tr}[
(j_\tau)^2)]} \, j_\tau + \frac{b(t)}{\text{tr}[ (j')^2 ]} \, j'
\end{equation}
with the time-dependent parts
\begin{equation}
a(t)= \text{tr}[j_\tau \rho_\text{I}(t)] \, , \quad b(t)= \text{tr}[ j'
\rho_\text{I}(t)] \, .
\end{equation}
Using the initial condition $\rho(0)$ in Eq.\
(\ref{eq:initial_condition}), one finds
\begin{equation}
a(t) = \alpha \, \frac{\text{tr}[e^{-i H_0 t} j_\tau e^{i H_0 t} e^{-i H t} j
e^{i H t}]}{d}
\end{equation}
and, evaluated at $t=\tau$,
\begin{equation}
a(\tau) = \alpha \, C_S(\tau) \, .
\end{equation}
Similarly, one finds
\begin{equation}\label{eq:b(t)}
b(\tau) = \alpha \left[ C_M(\tau) - \frac{C_S(\tau)_{\varepsilon =
0}}{C_S(0)_{\varepsilon = 0}} \, C_S(\tau)\right] \, .
\end{equation}
Hence, $a(\tau)$ is directly connected to the desired standard current
autocorrelation function $C_S(t)$, while $b(\tau$) is also related to
the modified current autocorrelation
function $C_M(t)$.

For the projection in Eq.\ (\ref{eq:complicated_projection}), the second-order
TCL
description in Eq.\ (\ref{eq:TCL2}) leads to the rate equation
\begin{equation} \label{eq:rate_equation_2D}
\frac{\text{d}}{\text{d}t} \begin{pmatrix} a(t) \\ b(t) \end{pmatrix}
= -\varepsilon^2 \, \Gamma_2(t) \begin{pmatrix} a(t) \\ b(t) \end{pmatrix} \, ,
\end{equation}
which now becomes twodimensional. In particular, $a(t)$ and $b(t)$ are coupled
via a time-dependent rate matrix $\Gamma_2(t)$. Similarly as before, this rate
matrix is given as a time integral over an kernel,
\begin{equation} \label{eq:rate_2D}
\Gamma_2(t) = \int_0^t \text{d}t' \, K_2(t,t') \, ,
\end{equation}
where the matrix elements of the kernel are given by
\begin{equation} \label{eq:K11}
K_{2, (1,1)}(t,t') = \frac{\text{tr}(j_\tau [V_I(t), [V_I(t'),
j_\tau]] )}{\text{tr}([j_\tau]^2)} \, ,
\end{equation}
\begin{equation}
K_{2, (1,2)}(t,t') = \frac{\text{tr}(j_\tau [V_I(t), [V_I(t'),
j']] )}{\text{tr}([j']^2)} \, ,
\end{equation}
\begin{equation}
K_{2, (2,1)}(t,t') = \frac{\text{tr}(j' [V_I(t), [V_I(t'),
j_\tau]] )}{\text{tr}([j_\tau]^2)} \, ,
\end{equation}
\begin{equation} \label{eq:K22}
K_{2, (2,2)}(t,t') = \frac{\text{tr}(j' [V_I(t), [V_I(t'),
j']] )}{\text{tr}([j']^2)} \, .
\end{equation}
and we assume that $K_2(t,t') = K_2(t-t')$.

Because the time dependence of the kernel $K_2(t,t')$ is generated by the
unperturbed system $H_0$ only, it is in principle possible to carry out the
perturbation theory analytically. However, $H_0$ must be solved exactly,
which can be challenging despite the Bethe-Ansatz integrability of the
spin-$1/2$ XXZ chain. Therefore, we carry out the perturbation theory
numerically, using again the concept of dynamical quantum typicality
\cite{heitmann2020, jin2021}. In comparison to standard
applications, the present implementation turns out to be more complex. Details
on this implementation can be found in Appendix \ref{sec:details}.


\section{Results}\label{sec:results}

In the following, we turn to our numerical simulations for the current
autocorrelation function in the perturbed spin-$1/2$ XXZ chain, where we
cover two different choices of the reference system. In Sec.\
\ref{sec:results_noninteracting}, we start with the noninteracting
reference system and particularly connect to existing results in the
literature. In Sec.\ \ref{sec:results_interacting}, we move forward to the
interacting reference system, which is the central case in our work.


\subsection{Noninteracting reference system} \label{sec:results_noninteracting}

Let us start with the noninteracting reference system $H_0 = H (\Delta = \Delta'
= 0)$, where the perturbation $V$ consists of interactions between both,
nearest and next-nearest sites. The strength of these two interactions is
chosen to be the same, $\varepsilon = \Delta = \Delta'$.
Since in the noninteracting reference system the current is conserved,
$[j,H_0] = 0$, the dynamics of the current autocorrelation function is
induced solely by the perturbation.
The conservation of the current in the noninteracting reference system
also implies that the standard and modified current-current correlation
functions are identical,
\begin{equation}
C_S(t) = C_M(t) \, .
\end{equation}
Thus, we do not distinguish between them here.

In Fig.\ \ref{fig:Fig1} (c), we summarize the temporal decay of the current
autocorrelation function $C_S(t)$ for a fixed system size $L = 30$ and various
perturbation strengths $\varepsilon$. Here, we particularly compare
data from exact simulations on the basis of dynamical quantum typicality to
data from second-order TCL. The overall agreement between the data is convincing
for all values of $\varepsilon$ depicted. Although deviations become visible
for strong perturbations, the rough shape of the two decay curves is still
similar, which is remarkable in view of the second-order truncation of TCL. A
comparison for other system sizes can be found in Appendix
\ref{sec:finite_size}.

It is also instructive to show data for the second-order kernel $k_2(t)$ and
rate $\gamma_2(t)$. The kernel $k_2(t)$ in Fig.\ \ref{fig:Fig1} (a) decays
quickly and remains close to zero, until finite-size effects become relevant.
Consequently, the rate $\gamma_2(t)$  in Fig.\ \ref{fig:Fig1} (b) quickly
saturates at a constant plateau value, until the same finite-size effects set
in. While the plateau value does not change significantly with system size, the
plateau length increases gradually. Therefore, already on the basis
of data for system sizes $L \leq 30$, one can conclude on the thermodynamic
limit. And indeed, the plateau value for $L \leq 30$ is consistent with the one
for $L = 2000$ in Ref.\ \cite{steinigeweg2011}, where an analytical expression
has been derived. Such a derivation is feasible, because of the noninteracting
reference system.

Let us briefly comment on the functional form of the decay
curve of the current autocorrelation function. Due to the saturation of the
rate $\gamma_2(t)$ at a constant plateau value, this decay curve is exponential
for small $\varepsilon$, where the relevant time scale is long and the initial
increase of $\gamma_2(t)$ can be neglected. For large $\varepsilon$, however,
the relevant time scale becomes short and the initial increase of $\gamma_2(t)$
is relevant. At this time scale, the decay curve turns into a Gaussian, which
is at least consistent with numerical simulations in Fig.\ \ref{fig:Fig1} (c).
Nevertheless, deviations from a clean Gaussian are clearly visible. For a more
detailed discussion of the current decay for strong perturbations, see
\cite{steinigeweg2011}.

\begin{figure}[t]
\includegraphics[width=0.9\columnwidth]{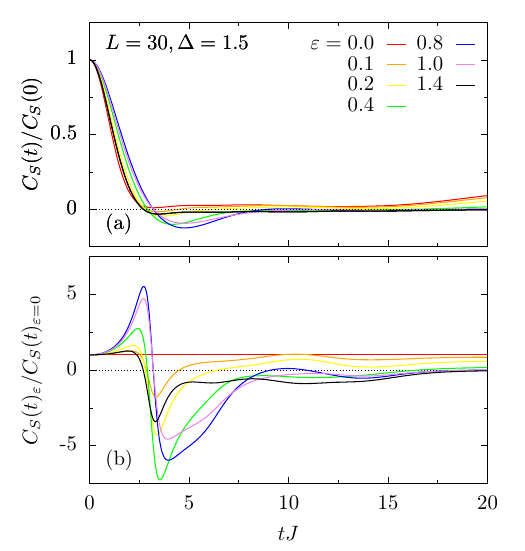}
\caption{Interacting reference system $H_0$ with $\Delta = 1.5$ and $\Delta' =
0$. (a) Exact dynamics of the autocorrelation function $C_S(t)$
for various perturbation strengths $\varepsilon$ and fixed system size $L=30$.
(b) Ratio between perturbed and unperturbed autocorrelation functions.}
\label{fig:Fig5}
\end{figure}

\begin{figure}[b]
\includegraphics[width=0.9\columnwidth]{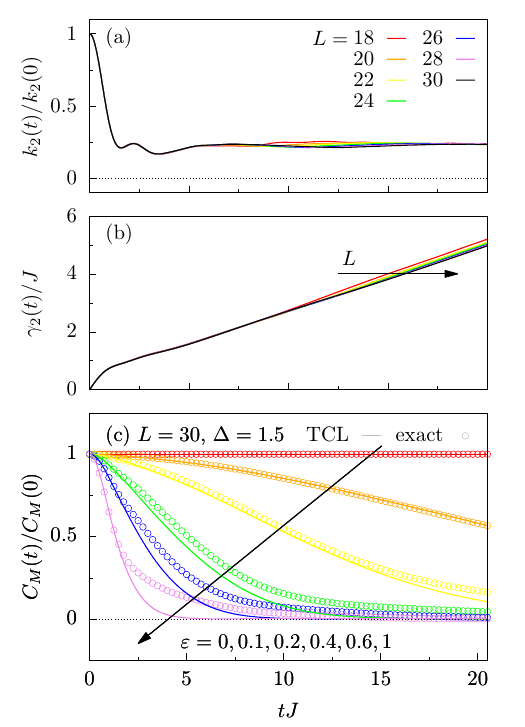}
\caption{Interacting reference system $H_0$ with $\Delta = 1.5$ and $\Delta' =
0$. (a) Second-order kernel $k_2(t)$ and (b) corresponding rate $\gamma_2(t)$,
both for various system sizes $L \leq 30$. (c) Resulting prediction for the
modified current autocorrelation function $C_M(t)$ and exact numerical results
for various perturbation strengths $\varepsilon$, both for a fixed system size
$L=30$.}
\label{fig:Fig6}
\end{figure}

\subsection{Interacting reference system} \label{sec:results_interacting}

Now, we are ready to come to the central part of our work, where we consider an
interacting reference system $H_0 = H(\Delta \neq 0, \Delta' =0)$. Because the
reference system now  includes also interactions between nearest sites,
the role of the perturbation $V$ is only played by interactions between
next-nearest sites. In the interacting reference system, the current is no
longer conserved, $[j,H_0] \neq 0$, and the standard and modified
current autocorrelation functions differ,
\begin{equation}
C_S(t) \neq C_M(t) \, .
\end{equation}
Here, we focus on a large anisotropy $\Delta = 8.0$, where this difference is
pronounced. Other values of the anisotropy are discussed in Appendix
\ref{sec:other_values}.

In Fig.\ \ref{fig:Fig2} (a), we first summarize the temporal decay of the
standard current autocorrelation function $C_S(t)$, as obtained from exact
numerical simulations. We do so for a fixed system size $L=30$ and various
perturbation strengths $\varepsilon$. While $C_S(t)$ is damped, it additionally
has oscillations, where frequencies and zero crossings depend on
$\varepsilon$. In particular, there is no obvious damping relation between the
perturbed and unperturbed dynamics. This observation becomes evident from
the ratio
\begin{equation}
C_S(t)_\varepsilon / C_S(t)_{\varepsilon = 0} \, ,
\end{equation}
as shown in Fig.\ \ref{fig:Fig2} (b). Consequently, the damping of the standard
current autocorrelation function $C_S(t)$ is highly nontrivial.

\begin{figure}[b]
\includegraphics[width=0.9\columnwidth]{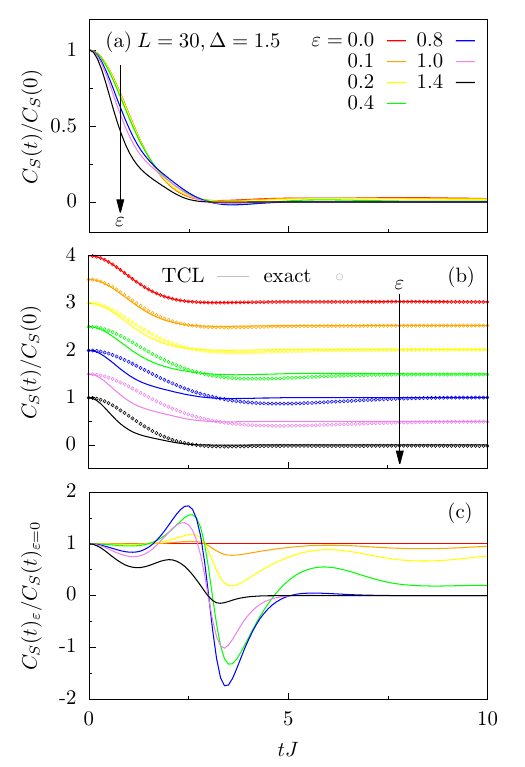}
\caption{Interacting reference system $H_0$ with $\Delta = 1.5$ and $\Delta' =
0$. (a) Second-order prediction for the dynamics of the standard current
autocorrelation function $C_S(t)$ for various perturbation strengths
$\varepsilon$ and a fixed system size $L=30$. (b) Comparison of (a) with the
exact dynamics, where data is shifted for better visibility. (c) Ratio between
perturbed and unperturbed
time evolution in (a).}
\label{fig:Fig7}
\end{figure}

In Fig.\ \ref{fig:Fig3} (c), we depict corresponding results for the temporal
decay of the modified current autocorrelation function $C_M(t)$. In clear
contrast to the standard $C_S(t)$ above, the modified $C_M(t)$ has a much
simpler relaxation behavior and is in favor of a simple damping relation. In
particular, exact data for $C_M(t)$ are in overall agreement
with data from second-order TCL. Still, the deviations in Fig.\ \ref{fig:Fig3}
(c) are larger than the ones in Fig.\ \ref{fig:Fig2} (c). 
However, it should be noted that the exact data for
$\Delta = 8.0$ are influenced by finite-size effects for times 
$t J \gtrsim {\cal O}(10)$~\cite{steinigeweg2014,karrasch2014}, even for 
a system size $L = 30$. At the same time, higher-order terms of the TCL
approach might also be needed to capture the dynamics
for large $\varepsilon$ accurately.

In Figs.\ \ref{fig:Fig3} (a) and (b), we further show the second-order kernel
$k_2(t)$ and rate $\gamma_2(t)$, respectively. Since $k_2(t)$ does not decay to
zero but saturates at a positive value, $\gamma_2(t)$ increases linearly, also
at long times. Consequently, the decay of the modified current autocorrelation
function $C_M(t)$ turns out to be Gaussian rather than exponential, even for
small perturbation strengths $\varepsilon$.

While the quite simple damping of the modified current autocorrelation function
$C_M(t)$ can be well understood by perturbation theory, a crucial question
remains: Can the highly nontrivial damping of the standard current
autocorrelation function $C_S(t)$ also be explained in this way? Or is this
damping a nonperturbative effect? To answer this question, we now replace the
projection in Eq.\ (\ref{eq:simple_projection}) by the one in Eq.\
(\ref{eq:complicated_projection}) and carry out the more involved second-order
TCL description. We show the so obtained results in Fig.\ \ref{fig:Fig4} (a) and
compare to exact data in Fig.\ \ref{fig:Fig4} (b). Apparently, the quality of
the agreement is as good as before. Moreover, as illustrated in Fig.\
\ref{fig:Fig4} (c), the perturbation theory is able to capture the nontrivial
features of the ratio between perturbed and unperturbed dynamics.


\section{Conclusions}\label{sec:conclusions}

We have discussed the impact of perturbations on the dynamics of quantum
many-body systems, with the focus on the magnetization current in the
spin-$1/2$ XXZ chain as a paradigmatic example for observable and model.
In particular, we have studied the damping of the current autocorrelation
function for two different choices of the reference system, where the first
one is noninteracting and second one is interacting. To this end, we have used
numerical simulations based on the concept of dynamical quantum typicality, in
order to obtain exact results for systems of comparatively large but still
finite size. We have complemented these results by further results from a
lowest-order perturbation theory on the basis of the TCL projection-operator
technique. For the noninteracting reference system, we have found an exponential
damping, which is consistent with other works. For the interacting reference
system, however, we have shown that the standard autocorrelation function is
damped in a nontrivial way and only a modified version of it is damped in a
simple manner. We have demonstrated that both, the nontrivial and simple damping
can be also understood on perturbative grounds. Our results are in agreement
with earlier findings for the particle current in the Hubbard chain
\cite{heitmann2021}, and they suggest that nontrivial damping occurs if the
unperturbed system already has a rich dynamical behavior.

Because the interacting reference system is already a complex problem despite
its Bethe-Ansatz integrability, we have carried out the lowest-order
perturbation theory on the basis of numerical simulations, which are rather
costly but still feasible for quite large system sizes. An analytical
calculation of this perturbation theory might be possible for other integrable
reference systems, which are less complex than the spin-$1/2$ XXZ chain.
Promising directions of future research include the incorporation of
higher-order corrections in the perturbation theory, the application to other
observables than currents, as well as the consideration of lower temperatures.


\subsection*{Acknowledgments}

We thank Jochen Gemmer for fruitful discussions.

This work has been funded by the DFG, under Grant No.\ 531128043, as well
as under Grant No.\ 397107022, No.\ 397067869, and No.\ 397082825 within DFG
Research Unit FOR 2692, under Grant No.\ 355031190.

Additionally, we greatly acknowledge computing time on the HPC3 at the
University of Osnabr\"uck, granted by the DFG, under Grant No.\ 456666331.

\subsection*{Data Availability}
The data that support the findings of this article are openly
available \cite{data}.

\appendix


\begin{figure}[b]
\includegraphics[width=0.9\columnwidth]{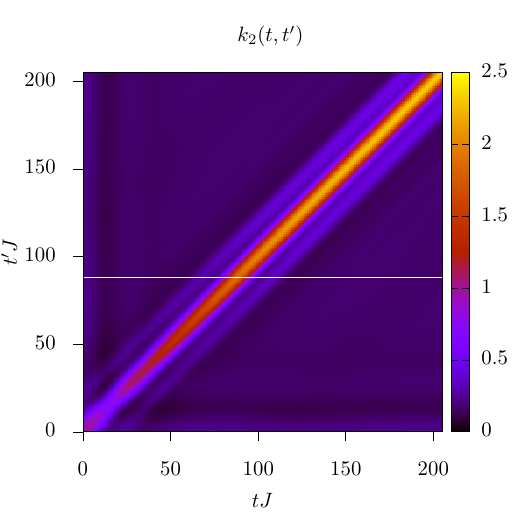}
\caption{Density plot of the second-order TCL kernel $k_2(t, t')$ in Eq.\
(\ref{eq:k2}) for an interacting reference $H_0$ with $\Delta = 1.5$ and
$\Delta' = 0$ and system size of $L=30$. A clear diagonal
structure is visible.}
\label{fig:Fig8}
\end{figure}

\section{Other values for the nearest-neighbor interaction}
\label{sec:other_values}

In the main text, we have focused on an interacting reference system with
an anisotropy $\Delta = 8.0$. Therefore, we redo the numerical calculations in
Figs.\ \ref{fig:Fig2}, \ref{fig:Fig3}, and \ref{fig:Fig4} for another
anisotropy $\Delta = 1.5$. As shown in Figs.\ \ref{fig:Fig5},
\ref{fig:Fig6}, and \ref{fig:Fig7}, the overall picture remains the same, which
indicates no substantial qualitative dependence on $\Delta$.


\section{Reduction of times} \label{sec:reduction}

As mentioned in the main text, the second-order TCL kernel $k_2(t,t')$ in Eq.\
(\ref{eq:k2}) generally depends on both time arguments $t$ and $t'$. While
$k_2(t,t') = k(t-t')$ holds for $[j, H_0] = 0$, this dependence on the mere
time difference $t-t'$ is an additional approximation
for $[j, H_0] \neq 0$, which we mainly use to simplify numerical simulations.
Hence, in Fig.\ \ref{fig:Fig8}, we support this approximation by showing that
$k_2(t,t')$ has a stripe structure around the diagonal $t = t'$.


\section{Finite-size analysis} \label{sec:finite_size}
\begin{figure}[t]
\includegraphics[width=0.9\columnwidth]{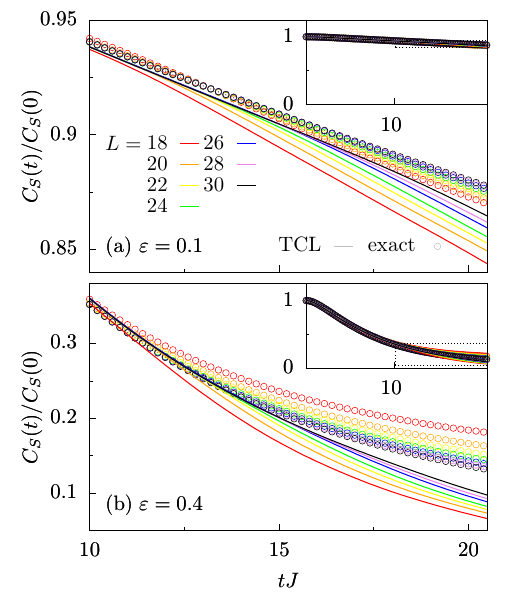}
\caption{Noninteracting reference system $H_0$ with $\Delta = 0$ and $\Delta' =
0$. Finite-size scaling of the current autocorrelation function $C_S(t)$ for
perturbation strengths (a) $\varepsilon=0.1$ and (b)
$\varepsilon=0.4$. Second-order TCL is compared to exact numerical simulations.
Insets: Different plot range. Here, also the plot region of the main panels is
indicated.}
\label{fig:Fig9}
\end{figure}

In the main text, we have compared the second-order TCL prediction to exact
numerical simulations for a fixed system size, e.g., $L = 30$. Thus, we redo the
comparison in Fig.\ \ref{fig:Fig1} (c) and summarize results for various $L$
and two different perturbation strengths $\varepsilon$ in Fig.\ \ref{fig:Fig9}.
As can been seen, the agreement improves with increasing $L$ and extends to
longer times.


\section{Numerical Details} \label{sec:details}

In all numerical simulations, we have used the concept of
dynamical quantum typicality. This concept has been used for the calculation of
standard \cite{heitmann2020, jin2021} and modified \cite{heitmann2021}
autocorrelation functions before. Therefore, we briefly describe the
application to the TCL projection-operator technique in
Sec.\ \ref{sec:TCL_extended_projection_and_standard_correlation_function}, in
particular to the matrix elements of second-order kernel $K_2(t-t')$ in Eqs.\
(\ref{eq:K11}) - (\ref{eq:K22}). 

\begin{figure}[t]
\includegraphics[width=0.9\columnwidth]{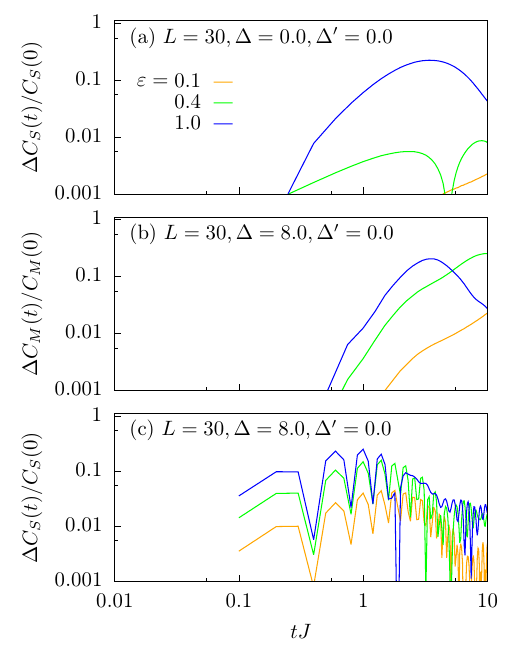}
\caption{Difference between TCL and exact data for
(a) the noninteracting reference system and (b), (c) the interacting reference
system, where the modified correlator
is depicted in (b) and the standard correlator is shown in (c). All data
are taken from Figs.\ \ref{fig:Fig1} (c), \ref{fig:Fig3} (c),
and \ref{fig:Fig4} (b).}
\label{fig:Fig11}
\end{figure}

Since the structure of the four matrix
elements is similar, we focus on, e.g.,
\begin{equation}
K_{2, (1,1)}(t,t') = \frac{\text{tr}(j_\tau [V_I(t), [V_I(t'),
j_\tau]] )}{\text{tr}([j_\tau]^2)}
\end{equation}
and its time-dependent numerator
\begin{equation}
\text{tr}(j_\tau [V_I(t), [V_I(t'), j_\tau]] ) \, ,
\end{equation}
because the time-independent denominator can be easily calculated
analytically. 
Executing the commutators, this numerator can be written as
\begin{eqnarray}
&& \! \text{tr}[ j_\tau V_\text{I}(t) V_\text{I}(t') j_\tau]
- \text{tr}[ j_\tau V_\text{I}(t) j_\tau V_\text{I}(t') \nonumber \\
&-& \! \text{tr}[V_\text{I}(t) j_\tau V_\text{I}(t') j_\tau]
+ \text{tr}[V_\text{I}(t) j_\tau j_\tau V_\text{I}(t')] \, ,
\end{eqnarray}
where the second and third term are identical, due to cyclic invariance of
the trace.
Since the structure of the terms is similar, we focus on, e.g.,
\begin{equation}
f(t,t') = \text{tr}[ j_\tau V_\text{I}(t) V_\text{I}(t') j_\tau]
\end{equation}
and assume that only the time difference matters,
\begin{equation}
f(t,t') = \text{tr}[ j_\tau V_\text{I}(t-t') V j_\tau] \, .
\end{equation}

\begin{figure}[b]

\includegraphics[width=0.9\columnwidth]{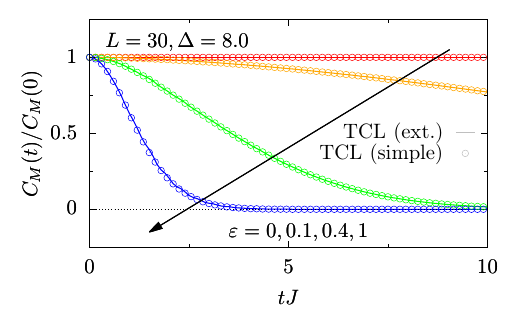}
\caption{Second-order TCL results for the modified current
autocorrelation function $C_M(t)$ for the interacting reference system $H_0$ 
with $\Delta=8.0$ and $\Delta'=0.0$, different perturbation strengths
$\varepsilon$, and system size $L=30$. Results from the simple and extended
projection are close to each other.}
\label{fig:Fig12}
\end{figure}

Now, using dynamical quantum typicality, we can replace the trace by an
expectation value of a Haar-random pure state,
\begin{equation}
f(t,t') = \langle \psi | j_\tau V_\text{I}(t-t') V j_\tau | \psi \rangle \, .
\end{equation}
By writing time-evolution operators explicitly and then rearranging, one
eventually gets
\begin{equation}
f(t,t') = \langle \psi_2(t) | V | \psi_1(t) \rangle
\end{equation}
with the two pure states
\begin{eqnarray}
&& | \psi_1(t) \rangle = e^{-i H_0 (t-t')} V e^{i H_0 \tau} j e^{-i H_0 \tau} |
\psi \rangle \, ,  \nonumber \\
&& | \psi_2(t) \rangle = e^{-i H_0 (t-t')} e^{i H_0 \tau} j e^{-i H_0 \tau} |
\psi \rangle \, .
\end{eqnarray}

The time evolution of these pure states can be obtained from
forward propagation on the basis of sparse-matrix techniques, as routinely done
in the context of dynamical quantum typicality \cite{heitmann2020, jin2021}.
However, in contrast to usual applications, more than one
time-evolution operator has to be treated. Moreover, the calculation has to be
done for (i) each term of a single matrix element, (ii) the in total four
matrix elements, and (iii) various values of the parameter $\tau$.
Thus, the overall calculation is rather costly but still feasible,
up to a quite large system size of, say, $L = 30$.

Once each matrix element of the second-order kernel $K_2(t-t')$ in Eqs.\
(\ref{eq:K11}) - (\ref{eq:K22}) is evaluated this way, the rate $\Gamma_2(t)$ in
Eq.\ (\ref{eq:rate_2D}) can be obtained by some numerical integration scheme.
Solving the differential equation (\ref{eq:rate_equation_2D}) as such is also
easy numerically and yields the functions $a(t)$ and $b(t)$, which have to be
evaluated at $t = \tau$ to get the current autocorrelation functions $C_S(\tau)$
and $C_M(\tau)$, respectively.


\section{Accuracy of the TCL approach} \label{sec:error analysis}
Here, we provide additional data for the accuracy of the TCL approach, by
calculating the deviations for the
standard correlation function,
\begin{equation}\label{eq:error}
\frac{\Delta C_S (t)}{C_S(0)}
=\bigg|\frac{C_S^{\text{exact}}(t)-C_S^{\text{TCL}}(t)}{C_S(0)}\bigg|\ , \\
\end{equation}
and for the modified correlation function,
\begin{equation}
\frac{\Delta C_M (t)}{C_M(0)}
=\bigg|\frac{C_M^{\text{exact}}(t)-C_M^{\text{TCL}}(t)}{C_M(0)}\bigg|\ .
\end{equation}
As depicted in Fig.~\ref{fig:Fig11}, a good accuracy is observed for
weak perturbations while deviations becomes visible for large perturbations,
especially if the reference system is interacting. However, as already
mentioned in the main text, it should be noted that the exact data for
${\Delta = 8.0}$ are influenced by finite-size effects for times
${t J \gtrsim {\cal O}(10)}$~\cite{steinigeweg2014,karrasch2014},
even for a system size $L = 30$. At the same time, higher-order terms of
the TCL approach might also be needed to capture the dynamics for
large $\varepsilon$ accurately.

In the main text, we have employed within the TCL approach
an extended projection to obtain the standard autocorrelation function.
For completeness, we employ it here to obtain the modified autocorrelation
function as well. As shown in Fig.~\ref{fig:Fig12}, the corresponding results
are nearly identical to those obtained using the simple projection.



\begin{thebibliography}{64}%
\makeatletter
\providecommand \@ifxundefined [1]{%
 \@ifx{#1\undefined}
}%
\providecommand \@ifnum [1]{%
 \ifnum #1\expandafter \@firstoftwo
 \else \expandafter \@secondoftwo
 \fi
}%
\providecommand \@ifx [1]{%
 \ifx #1\expandafter \@firstoftwo
 \else \expandafter \@secondoftwo
 \fi
}%
\providecommand \natexlab [1]{#1}%
\providecommand \enquote  [1]{``#1''}%
\providecommand \bibnamefont  [1]{#1}%
\providecommand \bibfnamefont [1]{#1}%
\providecommand \citenamefont [1]{#1}%
\providecommand \href@noop [0]{\@secondoftwo}%
\providecommand \href [0]{\begingroup \@sanitize@url \@href}%
\providecommand \@href[1]{\@@startlink{#1}\@@href}%
\providecommand \@@href[1]{\endgroup#1\@@endlink}%
\providecommand \@sanitize@url [0]{\catcode `\\12\catcode `\$12\catcode
  `\&12\catcode `\#12\catcode `\^12\catcode `\_12\catcode `\%12\relax}%
\providecommand \@@startlink[1]{}%
\providecommand \@@endlink[0]{}%
\providecommand \url  [0]{\begingroup\@sanitize@url \@url }%
\providecommand \@url [1]{\endgroup\@href {#1}{\urlprefix }}%
\providecommand \urlprefix  [0]{URL }%
\providecommand \Eprint [0]{\href }%
\providecommand \doibase [0]{http://dx.doi.org/}%
\providecommand \selectlanguage [0]{\@gobble}%
\providecommand \bibinfo  [0]{\@secondoftwo}%
\providecommand \bibfield  [0]{\@secondoftwo}%
\providecommand \translation [1]{[#1]}%
\providecommand \BibitemOpen [0]{}%
\providecommand \bibitemStop [0]{}%
\providecommand \bibitemNoStop [0]{.\EOS\space}%
\providecommand \EOS [0]{\spacefactor3000\relax}%
\providecommand \BibitemShut  [1]{\csname bibitem#1\endcsname}%
\let\auto@bib@innerbib\@empty
\bibitem [{\citenamefont {Polkovnikov}\ \emph {et~al.}(2011)\citenamefont
  {Polkovnikov}, \citenamefont {Sengupta}, \citenamefont {Silva},\ and\
  \citenamefont {Vengalattore}}]{polkovnikov2011}%
  \BibitemOpen
  \bibfield  {author} {\bibinfo {author} {\bibfnamefont {A.}~\bibnamefont
  {Polkovnikov}}, \bibinfo {author} {\bibfnamefont {K.}~\bibnamefont
  {Sengupta}}, \bibinfo {author} {\bibfnamefont {A.}~\bibnamefont {Silva}}, \
  and\ \bibinfo {author} {\bibfnamefont {M.}~\bibnamefont {Vengalattore}},\
  }\bibfield  {title} {\emph {\bibinfo {title} {Colloquium: Nonequilibrium
  dynamics of closed interacting quantum systems},\ }}\href
  {https://doi.org/10.1103/RevModPhys.83.863} {\bibfield  {journal} {\bibinfo
  {journal} {Rev. Mod. Phys.}\ }\textbf {\bibinfo {volume} {83}},\ \bibinfo
  {pages} {863} (\bibinfo {year} {2011})}\BibitemShut {NoStop}%
\bibitem [{\citenamefont {Eisert}\ \emph {et~al.}(2015)\citenamefont {Eisert},
  \citenamefont {Friesdorf},\ and\ \citenamefont {Gogolin}}]{eisert2015}%
  \BibitemOpen
  \bibfield  {author} {\bibinfo {author} {\bibfnamefont {J.}~\bibnamefont
  {Eisert}}, \bibinfo {author} {\bibfnamefont {M.}~\bibnamefont {Friesdorf}}, \
  and\ \bibinfo {author} {\bibfnamefont {C.}~\bibnamefont {Gogolin}},\
  }\bibfield  {title} {\emph {\bibinfo {title} {Quantum many-body systems out
  of equilibrium},\ }}\href {https://doi.org/10.1038/nphys3215} {\bibfield
  {journal} {\bibinfo  {journal} {Nat. Phys.}\ }\textbf {\bibinfo {volume}
  {11}},\ \bibinfo {pages} {124} (\bibinfo {year} {2015})}\BibitemShut
  {NoStop}%
\bibitem [{\citenamefont {Bloch}\ \emph {et~al.}(2008)\citenamefont {Bloch},
  \citenamefont {Dalibard},\ and\ \citenamefont {Zwerger}}]{bloch2008}%
  \BibitemOpen
  \bibfield  {author} {\bibinfo {author} {\bibfnamefont {I.}~\bibnamefont
  {Bloch}}, \bibinfo {author} {\bibfnamefont {J.}~\bibnamefont {Dalibard}}, \
  and\ \bibinfo {author} {\bibfnamefont {W.}~\bibnamefont {Zwerger}},\
  }\bibfield  {title} {\emph {\bibinfo {title} {Many-body physics with
  ultracold gases},\ }}\href {https://doi.org/10.1103/RevModPhys.80.885}
  {\bibfield  {journal} {\bibinfo  {journal} {Rev. Mod. Phys.}\ }\textbf
  {\bibinfo {volume} {80}},\ \bibinfo {pages} {885} (\bibinfo {year}
  {2008})}\BibitemShut {NoStop}%
\bibitem [{\citenamefont {D'Alessio}\ \emph {et~al.}(2016)\citenamefont
  {D'Alessio}, \citenamefont {Kafri}, \citenamefont {Polkovnikov},\ and\
  \citenamefont {Rigol}}]{dalessio2016}%
  \BibitemOpen
  \bibfield  {author} {\bibinfo {author} {\bibfnamefont {L.}~\bibnamefont
  {D'Alessio}}, \bibinfo {author} {\bibfnamefont {Y.}~\bibnamefont {Kafri}},
  \bibinfo {author} {\bibfnamefont {A.}~\bibnamefont {Polkovnikov}}, \ and\
  \bibinfo {author} {\bibfnamefont {M.}~\bibnamefont {Rigol}},\ }\bibfield
  {title} {\emph {\bibinfo {title} {From quantum chaos and eigenstate
  thermalization to statistical mechanics and thermodynamics},\ }}\href
  {https://doi.org/10.1080/00018732.2016.1198134} {\bibfield  {journal}
  {\bibinfo  {journal} {Adv. Phys.}\ }\textbf {\bibinfo {volume} {65}},\
  \bibinfo {pages} {239} (\bibinfo {year} {2016})}\BibitemShut {NoStop}%
\bibitem [{\citenamefont {Borgonovi}\ \emph {et~al.}(2016)\citenamefont
  {Borgonovi}, \citenamefont {Izrailev}, \citenamefont {Santos},\ and\
  \citenamefont {Zelevinsky}}]{borgonovi2016}%
  \BibitemOpen
  \bibfield  {author} {\bibinfo {author} {\bibfnamefont {F.}~\bibnamefont
  {Borgonovi}}, \bibinfo {author} {\bibfnamefont {F.}~\bibnamefont {Izrailev}},
  \bibinfo {author} {\bibfnamefont {L.}~\bibnamefont {Santos}}, \ and\ \bibinfo
  {author} {\bibfnamefont {V.}~\bibnamefont {Zelevinsky}},\ }\bibfield  {title}
  {\emph {\bibinfo {title} {Quantum chaos and thermalization in isolated
  systems of interacting particles},\ }}\href
  {https://doi.org/10.1016/j.physrep.2016.02.005} {\bibfield  {journal}
  {\bibinfo  {journal} {Phys. Rep.}\ }\textbf {\bibinfo {volume} {626}},\
  \bibinfo {pages} {1} (\bibinfo {year} {2016})}\BibitemShut {NoStop}%
\bibitem [{\citenamefont {Abanin}\ \emph {et~al.}(2019)\citenamefont {Abanin},
  \citenamefont {Altman}, \citenamefont {Bloch},\ and\ \citenamefont
  {Serbyn}}]{abanin2019}%
  \BibitemOpen
  \bibfield  {author} {\bibinfo {author} {\bibfnamefont {D.~A.}\ \bibnamefont
  {Abanin}}, \bibinfo {author} {\bibfnamefont {E.}~\bibnamefont {Altman}},
  \bibinfo {author} {\bibfnamefont {I.}~\bibnamefont {Bloch}}, \ and\ \bibinfo
  {author} {\bibfnamefont {M.}~\bibnamefont {Serbyn}},\ }\bibfield  {title}
  {\emph {\bibinfo {title} {Colloquium: Many-body localization, thermalization,
  and entanglement},\ }}\href {https://doi.org/10.1103/RevModPhys.91.021001}
  {\bibfield  {journal} {\bibinfo  {journal} {Rev. Mod. Phys.}\ }\textbf
  {\bibinfo {volume} {91}},\ \bibinfo {pages} {021001} (\bibinfo {year}
  {2019})}\BibitemShut {NoStop}%
\bibitem [{\citenamefont {Bertini}\ \emph {et~al.}(2021)\citenamefont
  {Bertini}, \citenamefont {Heidrich-Meisner}, \citenamefont {Karrasch},
  \citenamefont {Prosen}, \citenamefont {Steinigeweg},\ and\ \citenamefont
  {{\v{Z}}nidari{\v{c}}}}]{bertini2021}%
  \BibitemOpen
  \bibfield  {author} {\bibinfo {author} {\bibfnamefont {B.}~\bibnamefont
  {Bertini}}, \bibinfo {author} {\bibfnamefont {F.}~\bibnamefont
  {Heidrich-Meisner}}, \bibinfo {author} {\bibfnamefont {C.}~\bibnamefont
  {Karrasch}}, \bibinfo {author} {\bibfnamefont {T.}~\bibnamefont {Prosen}},
  \bibinfo {author} {\bibfnamefont {R.}~\bibnamefont {Steinigeweg}}, \ and\
  \bibinfo {author} {\bibfnamefont {M.}~\bibnamefont {{\v{Z}}nidari{\v{c}}}},\
  }\bibfield  {title} {\emph {\bibinfo {title} {Finite-temperature transport in
  one-dimensional quantum lattice models},\ }}\href
  {https://doi.org/10.1103/RevModPhys.93.025003} {\bibfield  {journal}
  {\bibinfo  {journal} {Rev. Mod. Phys.}\ }\textbf {\bibinfo {volume} {93}},\
  \bibinfo {pages} {025003} (\bibinfo {year} {2021})}\BibitemShut {NoStop}%
\bibitem [{\citenamefont {Landi}\ \emph {et~al.}(2022)\citenamefont {Landi},
  \citenamefont {Poletti},\ and\ \citenamefont {Schaller}}]{landi2022}%
  \BibitemOpen
  \bibfield  {author} {\bibinfo {author} {\bibfnamefont {G.~T.}\ \bibnamefont
  {Landi}}, \bibinfo {author} {\bibfnamefont {D.}~\bibnamefont {Poletti}}, \
  and\ \bibinfo {author} {\bibfnamefont {G.}~\bibnamefont {Schaller}},\
  }\bibfield  {title} {\emph {\bibinfo {title} {Nonequilibrium boundary-driven
  quantum systems: Models, methods, and properties},\ }}\href
  {https://doi.org/10.1103/RevModPhys.94.045006} {\bibfield  {journal}
  {\bibinfo  {journal} {Rev. Mod. Phys.}\ }\textbf {\bibinfo {volume} {94}},\
  \bibinfo {pages} {045006} (\bibinfo {year} {2022})}\BibitemShut {NoStop}%
\bibitem [{\citenamefont {Schollw\"ock}(2005)}]{schollwoeck2005}%
  \BibitemOpen
  \bibfield  {author} {\bibinfo {author} {\bibfnamefont {U.}~\bibnamefont
  {Schollw\"ock}},\ }\bibfield  {title} {\emph {\bibinfo {title} {The
  density-matrix renormalization group},\ }}\href
  {https://doi.org/10.1103/RevModPhys.77.259} {\bibfield  {journal} {\bibinfo
  {journal} {Rev. Mod. Phys.}\ }\textbf {\bibinfo {volume} {77}},\ \bibinfo
  {pages} {259} (\bibinfo {year} {2005})}\BibitemShut {NoStop}%
\bibitem [{\citenamefont {Schollw\"ock}(2011)}]{schollwoeck2011}%
  \BibitemOpen
  \bibfield  {author} {\bibinfo {author} {\bibfnamefont {U.}~\bibnamefont
  {Schollw\"ock}},\ }\bibfield  {title} {\emph {\bibinfo {title} {The
  density-matrix renormalization group in the age of matrix product states},\
  }}\href {https://doi.org/10.1016/j.aop.2010.09.012} {\bibfield  {journal}
  {\bibinfo  {journal} {Ann. Phys.}\ }\textbf {\bibinfo {volume} {326}},\
  \bibinfo {pages} {96} (\bibinfo {year} {2011})}\BibitemShut {NoStop}%
\bibitem [{\citenamefont {Weimer}\ \emph {et~al.}(2021)\citenamefont {Weimer},
  \citenamefont {Kshetrimayum},\ and\ \citenamefont {Or\'us}}]{weimer2021}%
  \BibitemOpen
  \bibfield  {author} {\bibinfo {author} {\bibfnamefont {H.}~\bibnamefont
  {Weimer}}, \bibinfo {author} {\bibfnamefont {A.}~\bibnamefont
  {Kshetrimayum}}, \ and\ \bibinfo {author} {\bibfnamefont {R.}~\bibnamefont
  {Or\'us}},\ }\bibfield  {title} {\emph {\bibinfo {title} {Simulation methods
  for open quantum many-body systems},\ }}\href
  {https://doi.org/RevModPhys.93.015008} {\bibfield  {journal} {\bibinfo
  {journal} {Rev. Mod. Phys.}\ }\textbf {\bibinfo {volume} {93}},\ \bibinfo
  {pages} {015008} (\bibinfo {year} {2021})}\BibitemShut {NoStop}%
\bibitem [{\citenamefont {Deutsch}(1991)}]{deutsch1991}%
  \BibitemOpen
  \bibfield  {author} {\bibinfo {author} {\bibfnamefont {J.~M.}\ \bibnamefont
  {Deutsch}},\ }\bibfield  {title} {\emph {\bibinfo {title} {Quantum
  statistical mechanics in a closed system},\ }}\href
  {https://doi.org/10.1103/PhysRevA.43.2046} {\bibfield  {journal} {\bibinfo
  {journal} {Phys. Rev. A}\ }\textbf {\bibinfo {volume} {43}},\ \bibinfo
  {pages} {2046} (\bibinfo {year} {1991})}\BibitemShut {NoStop}%
\bibitem [{\citenamefont {Srednicki}(1994)}]{srednicki1994}%
  \BibitemOpen
  \bibfield  {author} {\bibinfo {author} {\bibfnamefont {M.}~\bibnamefont
  {Srednicki}},\ }\bibfield  {title} {\emph {\bibinfo {title} {Chaos and
  quantum thermalization},\ }}\href {https://doi.org/10.1103/PhysRevE.50.888}
  {\bibfield  {journal} {\bibinfo  {journal} {Phys. Rev. E}\ }\textbf {\bibinfo
  {volume} {50}},\ \bibinfo {pages} {888} (\bibinfo {year} {1994})}\BibitemShut
  {NoStop}%
\bibitem [{\citenamefont {Rigol}\ \emph {et~al.}(2008)\citenamefont {Rigol},
  \citenamefont {Dunjko},\ and\ \citenamefont {Olshanii}}]{rigol2008}%
  \BibitemOpen
  \bibfield  {author} {\bibinfo {author} {\bibfnamefont {M.}~\bibnamefont
  {Rigol}}, \bibinfo {author} {\bibfnamefont {V.}~\bibnamefont {Dunjko}}, \
  and\ \bibinfo {author} {\bibfnamefont {M.}~\bibnamefont {Olshanii}},\
  }\bibfield  {title} {\emph {\bibinfo {title} {Thermalization and its
  mechanism for generic isolated quantum systems},\ }}\href
  {https://doi.org/10.1038/nature06838} {\bibfield  {journal} {\bibinfo
  {journal} {Nature}\ }\textbf {\bibinfo {volume} {452}},\ \bibinfo {pages}
  {854} (\bibinfo {year} {2008})}\BibitemShut {NoStop}%
\bibitem [{\citenamefont {Goldstein}\ \emph {et~al.}(2013)\citenamefont
  {Goldstein}, \citenamefont {Hara},\ and\ \citenamefont
  {Tasaki}}]{goldstein2013}%
  \BibitemOpen
  \bibfield  {author} {\bibinfo {author} {\bibfnamefont {S.}~\bibnamefont
  {Goldstein}}, \bibinfo {author} {\bibfnamefont {T.}~\bibnamefont {Hara}}, \
  and\ \bibinfo {author} {\bibfnamefont {H.}~\bibnamefont {Tasaki}},\
  }\bibfield  {title} {\emph {\bibinfo {title} {Time scales in the approach to
  equilibrium of macroscopic quantum systems},\ }}\href
  {https://doi.org/10.1103/PhysRevLett.111.140401} {\bibfield  {journal}
  {\bibinfo  {journal} {Phys. Rev. Lett.}\ }\textbf {\bibinfo {volume} {111}},\
  \bibinfo {pages} {140401} (\bibinfo {year} {2013})}\BibitemShut {NoStop}%
\bibitem [{\citenamefont {Reimann}(2016)}]{reimann2016}%
  \BibitemOpen
  \bibfield  {author} {\bibinfo {author} {\bibfnamefont {P.}~\bibnamefont
  {Reimann}},\ }\bibfield  {title} {\emph {\bibinfo {title} {Typical fast
  thermalization processes in closed many-body systems},\ }}\href
  {https://doi.org/10.1038/ncomms10821} {\bibfield  {journal} {\bibinfo
  {journal} {Nat. Commun.}\ }\textbf {\bibinfo {volume} {7}},\ \bibinfo {pages}
  {10821} (\bibinfo {year} {2016})}\BibitemShut {NoStop}%
\bibitem [{\citenamefont {Garc\'{\i}a-Pintos}\ \emph
  {et~al.}(2017)\citenamefont {Garc\'{\i}a-Pintos}, \citenamefont {Linden},
  \citenamefont {Malabarba}, \citenamefont {Short},\ and\ \citenamefont
  {Winter}}]{garciapintos2017}%
  \BibitemOpen
  \bibfield  {author} {\bibinfo {author} {\bibfnamefont {L.~P.}\ \bibnamefont
  {Garc\'{\i}a-Pintos}}, \bibinfo {author} {\bibfnamefont {N.}~\bibnamefont
  {Linden}}, \bibinfo {author} {\bibfnamefont {A.~S.~L.}\ \bibnamefont
  {Malabarba}}, \bibinfo {author} {\bibfnamefont {A.~J.}\ \bibnamefont
  {Short}}, \ and\ \bibinfo {author} {\bibfnamefont {A.}~\bibnamefont
  {Winter}},\ }\bibfield  {title} {\emph {\bibinfo {title} {Equilibration time
  scales of physically relevant observables},\ }}\href
  {https://doi.org/10.1103/PhysRevX.7.031027} {\bibfield  {journal} {\bibinfo
  {journal} {Phys. Rev. X}\ }\textbf {\bibinfo {volume} {7}},\ \bibinfo {pages}
  {031027} (\bibinfo {year} {2017})}\BibitemShut {NoStop}%
\bibitem [{\citenamefont {Richter}\ \emph {et~al.}(2019)\citenamefont
  {Richter}, \citenamefont {Gemmer},\ and\ \citenamefont
  {Steinigeweg}}]{richter2019}%
  \BibitemOpen
  \bibfield  {author} {\bibinfo {author} {\bibfnamefont {J.}~\bibnamefont
  {Richter}}, \bibinfo {author} {\bibfnamefont {J.}~\bibnamefont {Gemmer}}, \
  and\ \bibinfo {author} {\bibfnamefont {R.}~\bibnamefont {Steinigeweg}},\
  }\bibfield  {title} {\emph {\bibinfo {title} {Impact of eigenstate
  thermalization on the route to equilibrium},\ }}\href
  {https://doi.org/10.1103/PhysRevE.99.050104} {\bibfield  {journal} {\bibinfo
  {journal} {Phys. Rev. E}\ }\textbf {\bibinfo {volume} {99}},\ \bibinfo
  {pages} {050104} (\bibinfo {year} {2019})}\BibitemShut {NoStop}%
\bibitem [{\citenamefont {Alhambra}\ \emph {et~al.}(2020)\citenamefont
  {Alhambra}, \citenamefont {Riddell},\ and\ \citenamefont
  {Garc\'{\i}a-Pintos}}]{alhambra2020}%
  \BibitemOpen
  \bibfield  {author} {\bibinfo {author} {\bibfnamefont {A.~M.}\ \bibnamefont
  {Alhambra}}, \bibinfo {author} {\bibfnamefont {J.}~\bibnamefont {Riddell}}, \
  and\ \bibinfo {author} {\bibfnamefont {L.~P.}\ \bibnamefont
  {Garc\'{\i}a-Pintos}},\ }\bibfield  {title} {\emph {\bibinfo {title} {Time
  evolution of correlation functions in quantum many-body systems},\ }}\href
  {https://doi.org/10.1103/PhysRevLett.124.110605} {\bibfield  {journal}
  {\bibinfo  {journal} {Phys. Rev. Lett.}\ }\textbf {\bibinfo {volume} {124}},\
  \bibinfo {pages} {110605} (\bibinfo {year} {2020})}\BibitemShut {NoStop}%
\bibitem [{\citenamefont {Lezama}\ \emph {et~al.}(2021)\citenamefont {Lezama},
  \citenamefont {Torres-Herrera}, \citenamefont {P\'erez-Bernal}, \citenamefont
  {Bar~Lev},\ and\ \citenamefont {Santos}}]{lezama2021}%
  \BibitemOpen
  \bibfield  {author} {\bibinfo {author} {\bibfnamefont {T.~L.~M.}\
  \bibnamefont {Lezama}}, \bibinfo {author} {\bibfnamefont {E.~J.}\
  \bibnamefont {Torres-Herrera}}, \bibinfo {author} {\bibfnamefont
  {F.}~\bibnamefont {P\'erez-Bernal}}, \bibinfo {author} {\bibfnamefont
  {Y.}~\bibnamefont {Bar~Lev}}, \ and\ \bibinfo {author} {\bibfnamefont
  {L.~F.}\ \bibnamefont {Santos}},\ }\bibfield  {title} {\emph {\bibinfo
  {title} {Equilibration time in many-body quantum systems},\ }}\href
  {https://doi.org/10.1103/PhysRevB.104.085117} {\bibfield  {journal} {\bibinfo
   {journal} {Phys. Rev. B}\ }\textbf {\bibinfo {volume} {104}},\ \bibinfo
  {pages} {085117} (\bibinfo {year} {2021})}\BibitemShut {NoStop}%
\bibitem [{\citenamefont {Hamazaki}(2022)}]{hamazaki2022}%
  \BibitemOpen
  \bibfield  {author} {\bibinfo {author} {\bibfnamefont {R.}~\bibnamefont
  {Hamazaki}},\ }\bibfield  {title} {\emph {\bibinfo {title} {Speed limits for
  macroscopic transitions},\ }}\href
  {https://doi.org/10.1103/PRXQuantum.3.020319} {\bibfield  {journal} {\bibinfo
   {journal} {PRX Quantum}\ }\textbf {\bibinfo {volume} {3}},\ \bibinfo {pages}
  {020319} (\bibinfo {year} {2022})}\BibitemShut {NoStop}%
\bibitem [{\citenamefont {Bartsch}\ \emph {et~al.}(2024)\citenamefont
  {Bartsch}, \citenamefont {Dymarsky}, \citenamefont {Lamann}, \citenamefont
  {Wang}, \citenamefont {Steinigeweg},\ and\ \citenamefont
  {Gemmer}}]{bartsch2024}%
  \BibitemOpen
  \bibfield  {author} {\bibinfo {author} {\bibfnamefont {C.}~\bibnamefont
  {Bartsch}}, \bibinfo {author} {\bibfnamefont {A.}~\bibnamefont {Dymarsky}},
  \bibinfo {author} {\bibfnamefont {M.~H.}\ \bibnamefont {Lamann}}, \bibinfo
  {author} {\bibfnamefont {J.}~\bibnamefont {Wang}}, \bibinfo {author}
  {\bibfnamefont {R.}~\bibnamefont {Steinigeweg}}, \ and\ \bibinfo {author}
  {\bibfnamefont {J.}~\bibnamefont {Gemmer}},\ }\bibfield  {title} {\emph
  {\bibinfo {title} {Estimation of equilibration time scales from nested
  fraction approximations},\ }}\href
  {https://doi.org/10.1103/PhysRevE.110.024126} {\bibfield  {journal} {\bibinfo
   {journal} {Phys. Rev. E}\ }\textbf {\bibinfo {volume} {110}},\ \bibinfo
  {pages} {024126} (\bibinfo {year} {2024})}\BibitemShut {NoStop}%
\bibitem [{\citenamefont {Wang}\ \emph {et~al.}(2024)\citenamefont {Wang},
  \citenamefont {Füllgraf},\ and\ \citenamefont {Gemmer}}]{wang2024}%
  \BibitemOpen
  \bibfield  {author} {\bibinfo {author} {\bibfnamefont {J.}~\bibnamefont
  {Wang}}, \bibinfo {author} {\bibfnamefont {M.}~\bibnamefont {Füllgraf}}, \
  and\ \bibinfo {author} {\bibfnamefont {J.}~\bibnamefont {Gemmer}},\
  }\bibfield  {title} {\emph {\bibinfo {title} {Estimate of equilibration times
  of quantum correlation functions in the thermodynamic limit based on
  {L}anczos coefficients},\ }}\href {https://doi.org/10.48550/arXiv.2412.15932}
  {\bibfield  {journal} {\bibinfo  {journal} {arXiv:2412.15932}\ } (\bibinfo
  {year} {2024})}\BibitemShut {NoStop}%
\bibitem [{\citenamefont {Eckseler}\ and\ \citenamefont
  {Schnack}(2025)}]{eckseler2025}%
  \BibitemOpen
  \bibfield  {author} {\bibinfo {author} {\bibfnamefont {J.}~\bibnamefont
  {Eckseler}}\ and\ \bibinfo {author} {\bibfnamefont {J.}~\bibnamefont
  {Schnack}},\ }\bibfield  {title} {\emph {\bibinfo {title} {Permanent
  oscillations and solitary wave behavior in flatband {H}eisenberg quantum spin
  systems},\ }}\href {https://doi.org/10.1103/PhysRevResearch.7.013178}
  {\bibfield  {journal} {\bibinfo  {journal} {Phys. Rev. Res.}\ }\textbf
  {\bibinfo {volume} {7}},\ \bibinfo {pages} {013178} (\bibinfo {year}
  {2025})}\BibitemShut {NoStop}%
\bibitem [{\citenamefont {Zotos}(2004)}]{zotos2004}%
  \BibitemOpen
  \bibfield  {author} {\bibinfo {author} {\bibfnamefont {X.}~\bibnamefont
  {Zotos}},\ }\bibfield  {title} {\emph {\bibinfo {title} {High temperature
  thermal conductivity of two-leg spin-$1/2$ ladders},\ }}\href
  {https://doi.org/10.1103/PhysRevLett.92.067202} {\bibfield  {journal}
  {\bibinfo  {journal} {Phys. Rev. Lett.}\ }\textbf {\bibinfo {volume} {92}},\
  \bibinfo {pages} {067202} (\bibinfo {year} {2004})}\BibitemShut {NoStop}%
\bibitem [{\citenamefont {Jung}\ \emph {et~al.}(2006)\citenamefont {Jung},
  \citenamefont {Helmes},\ and\ \citenamefont {Rosch}}]{jung2006}%
  \BibitemOpen
  \bibfield  {author} {\bibinfo {author} {\bibfnamefont {P.}~\bibnamefont
  {Jung}}, \bibinfo {author} {\bibfnamefont {R.~W.}\ \bibnamefont {Helmes}}, \
  and\ \bibinfo {author} {\bibfnamefont {A.}~\bibnamefont {Rosch}},\ }\bibfield
   {title} {\emph {\bibinfo {title} {Transport in almost integrable models:
  Perturbed {H}eisenberg chains},\ }}\href
  {https://doi.org/10.1103/PhysRevLett.96.067202} {\bibfield  {journal}
  {\bibinfo  {journal} {Phys. Rev. Lett.}\ }\textbf {\bibinfo {volume} {96}},\
  \bibinfo {pages} {067202} (\bibinfo {year} {2006})}\BibitemShut {NoStop}%
\bibitem [{\citenamefont {Jung}\ and\ \citenamefont {Rosch}(2007)}]{jung2007}%
  \BibitemOpen
  \bibfield  {author} {\bibinfo {author} {\bibfnamefont {P.}~\bibnamefont
  {Jung}}\ and\ \bibinfo {author} {\bibfnamefont {A.}~\bibnamefont {Rosch}},\
  }\bibfield  {title} {\emph {\bibinfo {title} {Spin conductivity in almost
  integrable spin chains},\ }}\href
  {https://doi.org/10.1103/PhysRevB.76.245108} {\bibfield  {journal} {\bibinfo
  {journal} {Phys. Rev. B}\ }\textbf {\bibinfo {volume} {76}},\ \bibinfo
  {pages} {245108} (\bibinfo {year} {2007})}\BibitemShut {NoStop}%
\bibitem [{\citenamefont {Steinigeweg}\ \emph {et~al.}(2016)\citenamefont
  {Steinigeweg}, \citenamefont {Herbrych}, \citenamefont {Zotos},\ and\
  \citenamefont {Brenig}}]{steinigeweg2016}%
  \BibitemOpen
  \bibfield  {author} {\bibinfo {author} {\bibfnamefont {R.}~\bibnamefont
  {Steinigeweg}}, \bibinfo {author} {\bibfnamefont {J.}~\bibnamefont
  {Herbrych}}, \bibinfo {author} {\bibfnamefont {X.}~\bibnamefont {Zotos}}, \
  and\ \bibinfo {author} {\bibfnamefont {W.}~\bibnamefont {Brenig}},\
  }\bibfield  {title} {\emph {\bibinfo {title} {Heat conductivity of the
  {H}eisenberg spin-$1/2$ ladder: From weak to strong breaking of
  integrability},\ }}\href {https://doi.org/10.1103/PhysRevLett.116.017202}
  {\bibfield  {journal} {\bibinfo  {journal} {Phys. Rev. Lett.}\ }\textbf
  {\bibinfo {volume} {116}},\ \bibinfo {pages} {017202} (\bibinfo {year}
  {2016})}\BibitemShut {NoStop}%
\bibitem [{\citenamefont {De~Nardis}\ \emph {et~al.}(2021)\citenamefont
  {De~Nardis}, \citenamefont {Gopalakrishnan}, \citenamefont {Vasseur},\ and\
  \citenamefont {Ware}}]{denardis2021}%
  \BibitemOpen
  \bibfield  {author} {\bibinfo {author} {\bibfnamefont {J.}~\bibnamefont
  {De~Nardis}}, \bibinfo {author} {\bibfnamefont {S.}~\bibnamefont
  {Gopalakrishnan}}, \bibinfo {author} {\bibfnamefont {R.}~\bibnamefont
  {Vasseur}}, \ and\ \bibinfo {author} {\bibfnamefont {B.}~\bibnamefont
  {Ware}},\ }\bibfield  {title} {\emph {\bibinfo {title} {Stability of
  superdiffusion in nearly integrable spin chains},\ }}\href
  {https://doi.org/10.1103/PhysRevLett.127.057201} {\bibfield  {journal}
  {\bibinfo  {journal} {Phys. Rev. Lett.}\ }\textbf {\bibinfo {volume} {127}},\
  \bibinfo {pages} {057201} (\bibinfo {year} {2021})}\BibitemShut {NoStop}%
\bibitem [{\citenamefont {Mallayya}\ and\ \citenamefont
  {Rigol}(2021)}]{mallayya2021}%
  \BibitemOpen
  \bibfield  {author} {\bibinfo {author} {\bibfnamefont {K.}~\bibnamefont
  {Mallayya}}\ and\ \bibinfo {author} {\bibfnamefont {M.}~\bibnamefont
  {Rigol}},\ }\bibfield  {title} {\emph {\bibinfo {title} {Prethermalization,
  thermalization, and {F}ermi's golden rule in quantum many-body systems},\
  }}\href {https://doi.org/10.1103/PhysRevB.104.184302} {\bibfield  {journal}
  {\bibinfo  {journal} {Phys. Rev. B}\ }\textbf {\bibinfo {volume} {104}},\
  \bibinfo {pages} {184302} (\bibinfo {year} {2021})}\BibitemShut {NoStop}%
\bibitem [{\citenamefont {Roy}\ \emph {et~al.}(2023)\citenamefont {Roy},
  \citenamefont {Dhar}, \citenamefont {Spohn},\ and\ \citenamefont
  {Kulkarni}}]{roy2023}%
  \BibitemOpen
  \bibfield  {author} {\bibinfo {author} {\bibfnamefont {D.}~\bibnamefont
  {Roy}}, \bibinfo {author} {\bibfnamefont {A.}~\bibnamefont {Dhar}}, \bibinfo
  {author} {\bibfnamefont {H.}~\bibnamefont {Spohn}}, \ and\ \bibinfo {author}
  {\bibfnamefont {M.}~\bibnamefont {Kulkarni}},\ }\bibfield  {title} {\emph
  {\bibinfo {title} {Robustness of {K}ardar-{P}arisi-{Z}hang scaling in a
  classical integrable spin chain with broken integrability},\ }}\href
  {https://doi.org/10.1103/PhysRevB.107.L100413} {\bibfield  {journal}
  {\bibinfo  {journal} {Phys. Rev. B}\ }\textbf {\bibinfo {volume} {107}},\
  \bibinfo {pages} {L100413} (\bibinfo {year} {2023})}\BibitemShut {NoStop}%
\bibitem [{\citenamefont {Nandy}\ \emph {et~al.}(2023)\citenamefont {Nandy},
  \citenamefont {Lenar\ifmmode \check{c}\else
  \v{c}\fi{}i\ifmmode~\check{c}\else \v{c}\fi{}}, \citenamefont {Ilievski},
  \citenamefont {Mierzejewski}, \citenamefont {Herbrych},\ and\ \citenamefont
  {Prelov\ifmmode~\check{s}\else \v{s}\fi{}ek}}]{nandy2023}%
  \BibitemOpen
  \bibfield  {author} {\bibinfo {author} {\bibfnamefont {S.}~\bibnamefont
  {Nandy}}, \bibinfo {author} {\bibfnamefont {Z.}~\bibnamefont {Lenar\ifmmode
  \check{c}\else \v{c}\fi{}i\ifmmode~\check{c}\else \v{c}\fi{}}}, \bibinfo
  {author} {\bibfnamefont {E.}~\bibnamefont {Ilievski}}, \bibinfo {author}
  {\bibfnamefont {M.}~\bibnamefont {Mierzejewski}}, \bibinfo {author}
  {\bibfnamefont {J.}~\bibnamefont {Herbrych}}, \ and\ \bibinfo {author}
  {\bibfnamefont {P.}~\bibnamefont {Prelov\ifmmode~\check{s}\else
  \v{s}\fi{}ek}},\ }\bibfield  {title} {\emph {\bibinfo {title} {Spin diffusion
  in a perturbed isotropic {H}eisenberg spin chain},\ }}\href
  {https://doi.org/10.1103/PhysRevB.108.L081115} {\bibfield  {journal}
  {\bibinfo  {journal} {Phys. Rev. B}\ }\textbf {\bibinfo {volume} {108}},\
  \bibinfo {pages} {L081115} (\bibinfo {year} {2023})}\BibitemShut {NoStop}%
\bibitem [{\citenamefont {Gopalakrishnan}\ and\ \citenamefont
  {Vasseur}(2024)}]{gopalakrishnan2024}%
  \BibitemOpen
  \bibfield  {author} {\bibinfo {author} {\bibfnamefont {S.}~\bibnamefont
  {Gopalakrishnan}}\ and\ \bibinfo {author} {\bibfnamefont {R.}~\bibnamefont
  {Vasseur}},\ }\bibfield  {title} {\emph {\bibinfo {title} {Superdiffusion
  from nonabelian symmetries in nearly integrable systems},\ }}\href
  {https://doi.org/10.1146/annurev-conmatphys-032922-110710} {\bibfield
  {journal} {\bibinfo  {journal} {Annu. Rev. Condens. Matter Phys.}\ }\textbf
  {\bibinfo {volume} {15}},\ \bibinfo {pages} {159} (\bibinfo {year}
  {2024})}\BibitemShut {NoStop}%
\bibitem [{\citenamefont {Chen}\ \emph {et~al.}(2025)\citenamefont {Chen},
  \citenamefont {Huang}, \citenamefont {Ji}, \citenamefont {Schumacher},
  \citenamefont {Tsidilkovski}, \citenamefont {Schuckert}, \citenamefont
  {Assump\c{c}\~{a}o},\ and\ \citenamefont {Navon}}]{chen2025}%
  \BibitemOpen
  \bibfield  {author} {\bibinfo {author} {\bibfnamefont {J.}~\bibnamefont
  {Chen}}, \bibinfo {author} {\bibfnamefont {S.}~\bibnamefont {Huang}},
  \bibinfo {author} {\bibfnamefont {Y.}~\bibnamefont {Ji}}, \bibinfo {author}
  {\bibfnamefont {G.~L.}\ \bibnamefont {Schumacher}}, \bibinfo {author}
  {\bibfnamefont {A.}~\bibnamefont {Tsidilkovski}}, \bibinfo {author}
  {\bibfnamefont {A.}~\bibnamefont {Schuckert}}, \bibinfo {author}
  {\bibfnamefont {G.~G.~T.}\ \bibnamefont {Assump\c{c}\~{a}o}}, \ and\ \bibinfo
  {author} {\bibfnamefont {N.}~\bibnamefont {Navon}},\ }\bibfield  {title}
  {\emph {\bibinfo {title} {Emergence of {F}ermi's golden rule in the probing
  of a quantum many-body system},\ }}\href
  {https://doi.org/10.48550/arXiv.2502.14867} {\bibfield  {journal} {\bibinfo
  {journal} {arXiv:2502.14867}\ } (\bibinfo {year} {2025})}\BibitemShut
  {NoStop}%
\bibitem [{\citenamefont {Dabelow}\ and\ \citenamefont
  {Reimann}(2020)}]{dabelow2020}%
  \BibitemOpen
  \bibfield  {author} {\bibinfo {author} {\bibfnamefont {L.}~\bibnamefont
  {Dabelow}}\ and\ \bibinfo {author} {\bibfnamefont {P.}~\bibnamefont
  {Reimann}},\ }\bibfield  {title} {\emph {\bibinfo {title} {Relaxation theory
  for perturbed many-body quantum systems versus numerics and experiment},\
  }}\href {https://doi.org/10.1103/PhysRevLett.124.120602} {\bibfield
  {journal} {\bibinfo  {journal} {Phys. Rev. Lett.}\ }\textbf {\bibinfo
  {volume} {124}},\ \bibinfo {pages} {120602} (\bibinfo {year}
  {2020})}\BibitemShut {NoStop}%
\bibitem [{\citenamefont {Dabelow}\ and\ \citenamefont
  {Reimann}(2021)}]{dabelow2021}%
  \BibitemOpen
  \bibfield  {author} {\bibinfo {author} {\bibfnamefont {L.}~\bibnamefont
  {Dabelow}}\ and\ \bibinfo {author} {\bibfnamefont {P.}~\bibnamefont
  {Reimann}},\ }\bibfield  {title} {\emph {\bibinfo {title} {Typical relaxation
  of perturbed quantum many-body systems},\ }}\href
  {https://doi.org/10.1088/1742-5468/abd026} {\bibfield  {journal} {\bibinfo
  {journal} {J. Stat. Mech.: Theory Exp.}\ }\textbf {\bibinfo {volume}
  {2021}},\ \bibinfo {pages} {013106} (\bibinfo {year} {2021})}\BibitemShut
  {NoStop}%
\bibitem [{\citenamefont {Richter}\ \emph {et~al.}(2020)\citenamefont
  {Richter}, \citenamefont {Jin}, \citenamefont {Knipschild}, \citenamefont
  {De~Raedt}, \citenamefont {Michielsen}, \citenamefont {Gemmer},\ and\
  \citenamefont {Steinigeweg}}]{richter2020}%
  \BibitemOpen
  \bibfield  {author} {\bibinfo {author} {\bibfnamefont {J.}~\bibnamefont
  {Richter}}, \bibinfo {author} {\bibfnamefont {F.}~\bibnamefont {Jin}},
  \bibinfo {author} {\bibfnamefont {L.}~\bibnamefont {Knipschild}}, \bibinfo
  {author} {\bibfnamefont {H.}~\bibnamefont {De~Raedt}}, \bibinfo {author}
  {\bibfnamefont {K.}~\bibnamefont {Michielsen}}, \bibinfo {author}
  {\bibfnamefont {J.}~\bibnamefont {Gemmer}}, \ and\ \bibinfo {author}
  {\bibfnamefont {R.}~\bibnamefont {Steinigeweg}},\ }\bibfield  {title} {\emph
  {\bibinfo {title} {Exponential damping induced by random and realistic
  perturbations},\ }}\href {https://doi.org/10.1103/PhysRevE.101.062133}
  {\bibfield  {journal} {\bibinfo  {journal} {Phys. Rev. E}\ }\textbf {\bibinfo
  {volume} {101}},\ \bibinfo {pages} {062133} (\bibinfo {year}
  {2020})}\BibitemShut {NoStop}%
\bibitem [{\citenamefont {Heitmann}\ \emph {et~al.}(2021)\citenamefont
  {Heitmann}, \citenamefont {Richter}, \citenamefont {Gemmer},\ and\
  \citenamefont {Steinigeweg}}]{heitmann2021}%
  \BibitemOpen
  \bibfield  {author} {\bibinfo {author} {\bibfnamefont {T.}~\bibnamefont
  {Heitmann}}, \bibinfo {author} {\bibfnamefont {J.}~\bibnamefont {Richter}},
  \bibinfo {author} {\bibfnamefont {J.}~\bibnamefont {Gemmer}}, \ and\ \bibinfo
  {author} {\bibfnamefont {R.}~\bibnamefont {Steinigeweg}},\ }\bibfield
  {title} {\emph {\bibinfo {title} {Nontrivial damping of quantum many-body
  dynamics},\ }}\href {https://doi.org/10.1103/PhysRevE.104.054145} {\bibfield
  {journal} {\bibinfo  {journal} {Phys. Rev. E}\ }\textbf {\bibinfo {volume}
  {104}},\ \bibinfo {pages} {054145} (\bibinfo {year} {2021})}\BibitemShut
  {NoStop}%
\bibitem [{\citenamefont {Lamann}\ and\ \citenamefont
  {Gemmer}(2022)}]{lamann2022}%
  \BibitemOpen
  \bibfield  {author} {\bibinfo {author} {\bibfnamefont {M.~H.}\ \bibnamefont
  {Lamann}}\ and\ \bibinfo {author} {\bibfnamefont {J.}~\bibnamefont
  {Gemmer}},\ }\bibfield  {title} {\emph {\bibinfo {title} {Typical
  perturbation theory: Conditions, accuracy, and comparison with a mesoscopic
  case},\ }}\href {https://doi.org/10.1103/PhysRevE.106.054148} {\bibfield
  {journal} {\bibinfo  {journal} {Phys. Rev. E}\ }\textbf {\bibinfo {volume}
  {106}},\ \bibinfo {pages} {054148} (\bibinfo {year} {2022})}\BibitemShut
  {NoStop}%
\bibitem [{\citenamefont {Steinigeweg}\ and\ \citenamefont
  {Schnalle}(2010)}]{steinigeweg2010}%
  \BibitemOpen
  \bibfield  {author} {\bibinfo {author} {\bibfnamefont {R.}~\bibnamefont
  {Steinigeweg}}\ and\ \bibinfo {author} {\bibfnamefont {R.}~\bibnamefont
  {Schnalle}},\ }\bibfield  {title} {\emph {\bibinfo {title} {Projection
  operator approach to spin diffusion in the anisotropic {H}eisenberg chain at
  high temperatures},\ }}\href {https://doi.org/10.1103/PhysRevE.82.040103}
  {\bibfield  {journal} {\bibinfo  {journal} {Phys. Rev. E}\ }\textbf {\bibinfo
  {volume} {82}},\ \bibinfo {pages} {040103} (\bibinfo {year}
  {2010})}\BibitemShut {NoStop}%
\bibitem [{\citenamefont {Steinigeweg}(2011)}]{steinigeweg2011}%
  \BibitemOpen
  \bibfield  {author} {\bibinfo {author} {\bibfnamefont {R.}~\bibnamefont
  {Steinigeweg}},\ }\bibfield  {title} {\emph {\bibinfo {title} {Decay of
  currents for strong interactions},\ }}\href
  {https://doi.org/10.1103/PhysRevE.84.011136} {\bibfield  {journal} {\bibinfo
  {journal} {Phys. Rev. E}\ }\textbf {\bibinfo {volume} {84}},\ \bibinfo
  {pages} {011136} (\bibinfo {year} {2011})}\BibitemShut {NoStop}%
\bibitem [{\citenamefont {De~Nardis}\ \emph {et~al.}(2022)\citenamefont
  {De~Nardis}, \citenamefont {Gopalakrishnan}, \citenamefont {Vasseur},\ and\
  \citenamefont {Ware}}]{denardis2022}%
  \BibitemOpen
  \bibfield  {author} {\bibinfo {author} {\bibfnamefont {J.}~\bibnamefont
  {De~Nardis}}, \bibinfo {author} {\bibfnamefont {S.}~\bibnamefont
  {Gopalakrishnan}}, \bibinfo {author} {\bibfnamefont {R.}~\bibnamefont
  {Vasseur}}, \ and\ \bibinfo {author} {\bibfnamefont {B.}~\bibnamefont
  {Ware}},\ }\bibfield  {title} {\emph {\bibinfo {title} {Subdiffusive
  hydrodynamics of nearly integrable anisotropic spin chains},\ }}\href
  {https://doi.org/10.1073/pnas.2202823119} {\bibfield  {journal} {\bibinfo
  {journal} {Proc. Natl. Acad. Sci. USA}\ }\textbf {\bibinfo {volume} {119}}
  (\bibinfo {year} {2022})}\BibitemShut {NoStop}%
\bibitem [{\citenamefont {Prelov{\v{s}}ek}\ \emph {et~al.}(2022)\citenamefont
  {Prelov{\v{s}}ek}, \citenamefont {Nandy}, \citenamefont
  {Lenar{\v{c}}i{\v{c}}}, \citenamefont {Mierzejewski},\ and\ \citenamefont
  {Herbrych}}]{prelovsek2022}%
  \BibitemOpen
  \bibfield  {author} {\bibinfo {author} {\bibfnamefont {P.}~\bibnamefont
  {Prelov{\v{s}}ek}}, \bibinfo {author} {\bibfnamefont {S.}~\bibnamefont
  {Nandy}}, \bibinfo {author} {\bibfnamefont {Z.}~\bibnamefont
  {Lenar{\v{c}}i{\v{c}}}}, \bibinfo {author} {\bibfnamefont {M.}~\bibnamefont
  {Mierzejewski}}, \ and\ \bibinfo {author} {\bibfnamefont {J.}~\bibnamefont
  {Herbrych}},\ }\bibfield  {title} {\emph {\bibinfo {title} {{From
  dissipationless to normal diffusion in the easy-axis Heisenberg spin
  chain}},\ }}\href {https://doi.org/10.1103/PhysRevB.106.245104} {\bibfield
  {journal} {\bibinfo  {journal} {Phys. Rev. B}\ }\textbf {\bibinfo {volume}
  {106}},\ \bibinfo {pages} {245104} (\bibinfo {year} {2022})}\BibitemShut
  {NoStop}%
\bibitem [{\citenamefont {Kraft}\ \emph {et~al.}(2024)\citenamefont {Kraft},
  \citenamefont {Kempa}, \citenamefont {Wang}, \citenamefont {Nandy},\ and\
  \citenamefont {Steinigeweg}}]{kraft2024}%
  \BibitemOpen
  \bibfield  {author} {\bibinfo {author} {\bibfnamefont {M.}~\bibnamefont
  {Kraft}}, \bibinfo {author} {\bibfnamefont {M.}~\bibnamefont {Kempa}},
  \bibinfo {author} {\bibfnamefont {J.}~\bibnamefont {Wang}}, \bibinfo {author}
  {\bibfnamefont {S.}~\bibnamefont {Nandy}}, \ and\ \bibinfo {author}
  {\bibfnamefont {R.}~\bibnamefont {Steinigeweg}},\ }\bibfield  {title} {\emph
  {\bibinfo {title} {Scaling of diffusion constants in perturbed easy-axis
  {H}eisenberg spin chains},\ }}\href
  {https://doi.org/10.48550/arXiv.2410.22586} {\bibfield  {journal} {\bibinfo
  {journal} {arXiv:2410.22586}\ } (\bibinfo {year} {2024})}\BibitemShut
  {NoStop}%
\bibitem [{\citenamefont {Paw{\l}owski}\ \emph {et~al.}(2025)\citenamefont
  {Paw{\l}owski}, \citenamefont {Mierzejewski},\ and\ \citenamefont
  {Prelo\v{v}sek}}]{prelovsek2025}%
  \BibitemOpen
  \bibfield  {author} {\bibinfo {author} {\bibfnamefont {J.}~\bibnamefont
  {Paw{\l}owski}}, \bibinfo {author} {\bibfnamefont {M.}~\bibnamefont
  {Mierzejewski}}, \ and\ \bibinfo {author} {\bibfnamefont {P.}~\bibnamefont
  {Prelo\v{v}sek}},\ }\bibfield  {title} {\emph {\bibinfo {title} {Transport in
  integrable and perturbed easy-axis {H}eisenberg chain: Thouless approach},\
  }}\href {https://doi.org/10.48550/arXiv.2501.03735} {\bibfield  {journal}
  {\bibinfo  {journal} {arXiv:2501.03735}\ } (\bibinfo {year}
  {2025})}\BibitemShut {NoStop}%
\bibitem [{\citenamefont {Hams}\ and\ \citenamefont
  {De~Raedt}(2000)}]{hams2000}%
  \BibitemOpen
  \bibfield  {author} {\bibinfo {author} {\bibfnamefont {A.}~\bibnamefont
  {Hams}}\ and\ \bibinfo {author} {\bibfnamefont {H.}~\bibnamefont
  {De~Raedt}},\ }\bibfield  {title} {\emph {\bibinfo {title} {Fast algorithm
  for finding the eigenvalue distribution of very large matrices},\ }}\href
  {https://doi.org/10.1103/PhysRevE.62.4365} {\bibfield  {journal} {\bibinfo
  {journal} {Phys. Rev. E}\ }\textbf {\bibinfo {volume} {62}},\ \bibinfo
  {pages} {4365} (\bibinfo {year} {2000})}\BibitemShut {NoStop}%
\bibitem [{\citenamefont {Gemmer}\ and\ \citenamefont
  {Mahler}(2003)}]{gemmer2003}%
  \BibitemOpen
  \bibfield  {author} {\bibinfo {author} {\bibfnamefont {J.}~\bibnamefont
  {Gemmer}}\ and\ \bibinfo {author} {\bibfnamefont {G.}~\bibnamefont
  {Mahler}},\ }\bibfield  {title} {\emph {\bibinfo {title} {Distribution of
  local entropy in the {H}ilbert space of bi-partite quantum systems: origin of
  {J}aynes' principle},\ }}\href {https://doi.org/10.1140/epjb/e2003-00029-3}
  {\bibfield  {journal} {\bibinfo  {journal} {Eur. Phys. J. B}\ }\textbf
  {\bibinfo {volume} {31}},\ \bibinfo {pages} {249} (\bibinfo {year}
  {2003})}\BibitemShut {NoStop}%
\bibitem [{\citenamefont {Goldstein}\ \emph {et~al.}(2006)\citenamefont
  {Goldstein}, \citenamefont {Lebowitz}, \citenamefont {Tumulka},\ and\
  \citenamefont {Zangh\`{\i}}}]{goldstein2006}%
  \BibitemOpen
  \bibfield  {author} {\bibinfo {author} {\bibfnamefont {S.}~\bibnamefont
  {Goldstein}}, \bibinfo {author} {\bibfnamefont {J.~L.}\ \bibnamefont
  {Lebowitz}}, \bibinfo {author} {\bibfnamefont {R.}~\bibnamefont {Tumulka}}, \
  and\ \bibinfo {author} {\bibfnamefont {N.}~\bibnamefont {Zangh\`{\i}}},\
  }\bibfield  {title} {\emph {\bibinfo {title} {Canonical typicality},\ }}\href
  {https://doi.org/10.1103/PhysRevLett.96.050403} {\bibfield  {journal}
  {\bibinfo  {journal} {Phys. Rev. Lett.}\ }\textbf {\bibinfo {volume} {96}},\
  \bibinfo {pages} {050403} (\bibinfo {year} {2006})}\BibitemShut {NoStop}%
\bibitem [{\citenamefont {Popescu}\ \emph {et~al.}(2006)\citenamefont
  {Popescu}, \citenamefont {Short},\ and\ \citenamefont
  {Winter}}]{popescu2006}%
  \BibitemOpen
  \bibfield  {author} {\bibinfo {author} {\bibfnamefont {S.}~\bibnamefont
  {Popescu}}, \bibinfo {author} {\bibfnamefont {A.~J.}\ \bibnamefont {Short}},
  \ and\ \bibinfo {author} {\bibfnamefont {A.}~\bibnamefont {Winter}},\
  }\bibfield  {title} {\emph {\bibinfo {title} {Entanglement and the
  foundations of statistical mechanics},\ }}\href
  {https://doi.org/10.1038/nphys444} {\bibfield  {journal} {\bibinfo  {journal}
  {Nat. Phys.}\ }\textbf {\bibinfo {volume} {2}},\ \bibinfo {pages} {754}
  (\bibinfo {year} {2006})}\BibitemShut {NoStop}%
\bibitem [{\citenamefont {Reimann}(2007)}]{reimann2007}%
  \BibitemOpen
  \bibfield  {author} {\bibinfo {author} {\bibfnamefont {P.}~\bibnamefont
  {Reimann}},\ }\bibfield  {title} {\emph {\bibinfo {title} {Typicality for
  generalized microcanonical ensembles},\ }}\href
  {https://doi.org/10.1103/PhysRevLett.99.160404} {\bibfield  {journal}
  {\bibinfo  {journal} {Phys. Rev. Lett.}\ }\textbf {\bibinfo {volume} {99}},\
  \bibinfo {pages} {160404} (\bibinfo {year} {2007})}\BibitemShut {NoStop}%
\bibitem [{\citenamefont {Bartsch}\ and\ \citenamefont
  {Gemmer}(2009)}]{bartsch2009}%
  \BibitemOpen
  \bibfield  {author} {\bibinfo {author} {\bibfnamefont {C.}~\bibnamefont
  {Bartsch}}\ and\ \bibinfo {author} {\bibfnamefont {J.}~\bibnamefont
  {Gemmer}},\ }\bibfield  {title} {\emph {\bibinfo {title} {Dynamical
  typicality of quantum expectation values},\ }}\href
  {https://doi.org/10.1103/PhysRevLett.102.110403} {\bibfield  {journal}
  {\bibinfo  {journal} {Phys. Rev. Lett.}\ }\textbf {\bibinfo {volume} {102}},\
  \bibinfo {pages} {110403} (\bibinfo {year} {2009})}\BibitemShut {NoStop}%
\bibitem [{\citenamefont {White}(2009)}]{white2009}%
  \BibitemOpen
  \bibfield  {author} {\bibinfo {author} {\bibfnamefont {S.~R.}\ \bibnamefont
  {White}},\ }\bibfield  {title} {\emph {\bibinfo {title} {Minimally entangled
  typical quantum states at finite temperature},\ }}\href
  {https://doi.org/10.1103/PhysRevLett.102.190601} {\bibfield  {journal}
  {\bibinfo  {journal} {Phys. Rev. Lett.}\ }\textbf {\bibinfo {volume} {102}},\
  \bibinfo {pages} {190601} (\bibinfo {year} {2009})}\BibitemShut {NoStop}%
\bibitem [{\citenamefont {Bartsch}\ and\ \citenamefont
  {Gemmer}(2011)}]{bartsch2011}%
  \BibitemOpen
  \bibfield  {author} {\bibinfo {author} {\bibfnamefont {C.}~\bibnamefont
  {Bartsch}}\ and\ \bibinfo {author} {\bibfnamefont {J.}~\bibnamefont
  {Gemmer}},\ }\bibfield  {title} {\emph {\bibinfo {title} {Transient
  fluctuation theorem in closed quantum systems},\ }}\href
  {https://doi.org/10.1209/0295-5075/96/60008} {\bibfield  {journal} {\bibinfo
  {journal} {Europhys. Lett.}\ }\textbf {\bibinfo {volume} {96}},\ \bibinfo
  {pages} {60008} (\bibinfo {year} {2011})}\BibitemShut {NoStop}%
\bibitem [{\citenamefont {Sugiura}\ and\ \citenamefont
  {Shimizu}(2012)}]{sugiura2012}%
  \BibitemOpen
  \bibfield  {author} {\bibinfo {author} {\bibfnamefont {S.}~\bibnamefont
  {Sugiura}}\ and\ \bibinfo {author} {\bibfnamefont {A.}~\bibnamefont
  {Shimizu}},\ }\bibfield  {title} {\emph {\bibinfo {title} {Thermal pure
  quantum states at finite temperature},\ }}\href
  {https://doi.org/10.1103/PhysRevLett.108.240401} {\bibfield  {journal}
  {\bibinfo  {journal} {Phys. Rev. Lett.}\ }\textbf {\bibinfo {volume} {108}},\
  \bibinfo {pages} {240401} (\bibinfo {year} {2012})}\BibitemShut {NoStop}%
\bibitem [{\citenamefont {Elsayed}\ and\ \citenamefont
  {Fine}(2013)}]{elsayed2013}%
  \BibitemOpen
  \bibfield  {author} {\bibinfo {author} {\bibfnamefont {T.~A.}\ \bibnamefont
  {Elsayed}}\ and\ \bibinfo {author} {\bibfnamefont {B.~V.}\ \bibnamefont
  {Fine}},\ }\bibfield  {title} {\emph {\bibinfo {title} {Regression relation
  for pure quantum states and its implications for efficient computing},\
  }}\href {https://doi.org/10.1103/PhysRevLett.110.070404} {\bibfield
  {journal} {\bibinfo  {journal} {Phys. Rev. Lett.}\ }\textbf {\bibinfo
  {volume} {110}},\ \bibinfo {pages} {070404} (\bibinfo {year}
  {2013})}\BibitemShut {NoStop}%
\bibitem [{\citenamefont {Steinigeweg}\ \emph {et~al.}(2014)\citenamefont
  {Steinigeweg}, \citenamefont {Gemmer},\ and\ \citenamefont
  {Brenig}}]{steinigeweg2014}%
  \BibitemOpen
  \bibfield  {author} {\bibinfo {author} {\bibfnamefont {R.}~\bibnamefont
  {Steinigeweg}}, \bibinfo {author} {\bibfnamefont {J.}~\bibnamefont {Gemmer}},
  \ and\ \bibinfo {author} {\bibfnamefont {W.}~\bibnamefont {Brenig}},\
  }\bibfield  {title} {\emph {\bibinfo {title} {Spin-current autocorrelations
  from single pure-state propagation},\ }}\href
  {https://doi.org/10.1103/PhysRevLett.112.120601} {\bibfield  {journal}
  {\bibinfo  {journal} {Phys. Rev. Lett.}\ }\textbf {\bibinfo {volume} {112}},\
  \bibinfo {pages} {120601} (\bibinfo {year} {2014})}\BibitemShut {NoStop}%
\bibitem [{\citenamefont {Steinigeweg}\ \emph {et~al.}(2015)\citenamefont
  {Steinigeweg}, \citenamefont {Gemmer},\ and\ \citenamefont
  {Brenig}}]{steinigeweg2015}%
  \BibitemOpen
  \bibfield  {author} {\bibinfo {author} {\bibfnamefont {R.}~\bibnamefont
  {Steinigeweg}}, \bibinfo {author} {\bibfnamefont {J.}~\bibnamefont {Gemmer}},
  \ and\ \bibinfo {author} {\bibfnamefont {W.}~\bibnamefont {Brenig}},\
  }\bibfield  {title} {\emph {\bibinfo {title} {Spin and energy currents in
  integrable and nonintegrable spin-$\frac{1}{2}$ chains: A typicality approach
  to real-time autocorrelations},\ }}\href
  {https://doi.org/10.1103/PhysRevB.91.104404} {\bibfield  {journal} {\bibinfo
  {journal} {Phys. Rev. B}\ }\textbf {\bibinfo {volume} {91}},\ \bibinfo
  {pages} {104404} (\bibinfo {year} {2015})}\BibitemShut {NoStop}%
\bibitem [{\citenamefont {Heitmann}\ \emph {et~al.}(2020)\citenamefont
  {Heitmann}, \citenamefont {Richter}, \citenamefont {Schubert},\ and\
  \citenamefont {Steinigeweg}}]{heitmann2020}%
  \BibitemOpen
  \bibfield  {author} {\bibinfo {author} {\bibfnamefont {T.}~\bibnamefont
  {Heitmann}}, \bibinfo {author} {\bibfnamefont {J.}~\bibnamefont {Richter}},
  \bibinfo {author} {\bibfnamefont {D.}~\bibnamefont {Schubert}}, \ and\
  \bibinfo {author} {\bibfnamefont {R.}~\bibnamefont {Steinigeweg}},\
  }\bibfield  {title} {\emph {\bibinfo {title} {Selected applications of
  typicality to real-time dynamics of quantum many-body systems},\ }}\href
  {\doibase 10.1515/zna-2020-0010} {\bibfield  {journal} {\bibinfo  {journal}
  {Z. Naturforsch. A}\ }\textbf {\bibinfo {volume} {75}},\ \bibinfo {pages}
  {421} (\bibinfo {year} {2020})}\BibitemShut {NoStop}%
\bibitem [{\citenamefont {Jin}\ \emph {et~al.}(2021)\citenamefont {Jin},
  \citenamefont {Willsch}, \citenamefont {Willsch}, \citenamefont {Lagemann},
  \citenamefont {Michielsen},\ and\ \citenamefont {{De Raedt}}}]{jin2021}%
  \BibitemOpen
  \bibfield  {author} {\bibinfo {author} {\bibfnamefont {F.}~\bibnamefont
  {Jin}}, \bibinfo {author} {\bibfnamefont {D.}~\bibnamefont {Willsch}},
  \bibinfo {author} {\bibfnamefont {M.}~\bibnamefont {Willsch}}, \bibinfo
  {author} {\bibfnamefont {H.}~\bibnamefont {Lagemann}}, \bibinfo {author}
  {\bibfnamefont {K.}~\bibnamefont {Michielsen}}, \ and\ \bibinfo {author}
  {\bibfnamefont {H.}~\bibnamefont {{De Raedt}}},\ }\bibfield  {title} {\emph
  {\bibinfo {title} {{Random state technology}},\ }}\href
  {https://doi.org/10.7566/JPSJ.90.012001} {\bibfield  {journal} {\bibinfo
  {journal} {J. Phys. Soc. Jpn.}\ }\textbf {\bibinfo {volume} {90}},\ \bibinfo
  {pages} {012001} (\bibinfo {year} {2021})}\BibitemShut {NoStop}%
\bibitem [{\citenamefont {Mitri\'{c}}(2024)}]{mitric2024}%
  \BibitemOpen
  \bibfield  {author} {\bibinfo {author} {\bibfnamefont {P.}~\bibnamefont
  {Mitri\'{c}}},\ }\bibfield  {title} {\emph {\bibinfo {title} {Dynamical
  quantum typicality: A simple method for investigating transport properties
  applied to the {H}olstein model},\ }}\href
  {https://doi.org/10.48550/arXiv.2412.17436} {\bibfield  {journal} {\bibinfo
  {journal} {arXiv:2412.17436}\ } (\bibinfo {year} {2024})}\BibitemShut
  {NoStop}%
\bibitem [{\citenamefont {Chaturvedi}\ and\ \citenamefont
  {Shibata}(1979)}]{chaturvedi1979}%
  \BibitemOpen
  \bibfield  {author} {\bibinfo {author} {\bibfnamefont {S.}~\bibnamefont
  {Chaturvedi}}\ and\ \bibinfo {author} {\bibfnamefont {F.}~\bibnamefont
  {Shibata}},\ }\bibfield  {title} {\emph {\bibinfo {title}
  {Time-convolutionless projection operator formalism for elimination of fast
  variables. {A}pplications to {B}rownian motion},\ }}\href
  {https://doi.org/10.1007/bf01319852} {\bibfield  {journal} {\bibinfo
  {journal} {Z. Phys. B}\ }\textbf {\bibinfo {volume} {35}},\ \bibinfo {pages}
  {297} (\bibinfo {year} {1979})}\BibitemShut {NoStop}%
\bibitem [{\citenamefont {Breuer}\ and\ \citenamefont
  {Petruccione}(2007)}]{breuer2007}%
  \BibitemOpen
  \bibfield  {author} {\bibinfo {author} {\bibfnamefont {H.-P.}\ \bibnamefont
  {Breuer}}\ and\ \bibinfo {author} {\bibfnamefont {F.}~\bibnamefont
  {Petruccione}},\ }\href
  {https://doi.org/10.1093/acprof:oso/9780199213900.001.0001} {\emph {\bibinfo
  {title} {The theory of open quantum systems}}}\ (\bibinfo  {publisher}
  {Oxford University Press},\ \bibinfo {year} {2007})\BibitemShut {NoStop}%
\bibitem [{\citenamefont {Kubo}\ \emph {et~al.}(1991)\citenamefont {Kubo},
  \citenamefont {Toda},\ and\ \citenamefont {Hashisume}}]{kubo1991}%
  \BibitemOpen
  \bibfield  {author} {\bibinfo {author} {\bibfnamefont {R.}~\bibnamefont
  {Kubo}}, \bibinfo {author} {\bibfnamefont {M.}~\bibnamefont {Toda}}, \ and\
  \bibinfo {author} {\bibfnamefont {N.}~\bibnamefont {Hashisume}},\ }\href
  {https://doi.org/10.1007/978-3-642-58244-8} {\emph {\bibinfo {title}
  {Statistical physics II: Nonequilibrium statistical mechanics}}}\ (\bibinfo
  {publisher} {Springer, Berlin, Heidelberg},\ \bibinfo {year}
  {1991})\BibitemShut {NoStop}%
\bibitem [{\citenamefont {Karrasch}\ \emph {et~al.}(2014)\citenamefont
  {Karrasch}, \citenamefont {Moore},\ and\ \citenamefont
  {Heidrich-Meisner}}]{karrasch2014}%
  \BibitemOpen
  \bibfield  {author} {\bibinfo {author} {\bibfnamefont {C.}~\bibnamefont
  {Karrasch}}, \bibinfo {author} {\bibfnamefont {J.~E.}\ \bibnamefont {Moore}},
  \ and\ \bibinfo {author} {\bibfnamefont {F.}~\bibnamefont
  {Heidrich-Meisner}},\ }\bibfield  {title} {\emph {\bibinfo {title}
  {{Real-time and real-space spin and energy dynamics in one-dimensional
  spin-1/2 systems induced by local quantum quenches at finite temperatures}},\
  }}\href {https://doi.org/10.1103/PhysRevB.89.075139} {\bibfield  {journal}
  {\bibinfo  {journal} {Phys. Rev. B}\ }\textbf {\bibinfo {volume} {89}},\
  \bibinfo {pages} {075139} (\bibinfo {year} {2014})}\BibitemShut {NoStop}%
%
\bibitem{data} {M. Kempa, Nontrivial damping of
magnetization currents in perturbed spin chains [dataset], Zenodo (2025), doi:}
\href{https://doi.org/10.5281/zenodo.15834510}
  { {\bibinfo {title}
  {10.5281/zenodo.15834510}}}
\end{thebibliography}

%

\clearpage
\newpage

\end{document}